\documentclass[10pt,preprint2]{aastex}

\usepackage{txfonts}
\usepackage{ulem} 
\usepackage{natbib}
\usepackage{rotating}
\usepackage{subfigure}
\usepackage{multirow}
\usepackage{color}

\definecolor{ao}{rgb}{0.0, 0.5, 0.0}
\shorttitle{On the Solar Dynamo}
\shortauthors{Simoniello et al.}

\begin{document}

\title {A New Challenge to Solar Dynamo Models from Helioseismic Observations: \\ The Latitudinal Dependence of the Progression of the Solar Cycle}

%A New Challenge for Solar Dynamo Models: \\ The Helioseismic Signatures in the Progression of  the Solar Cycle}%: \\A Reflection of $\alpha\Omega$ Dynamo Wave}


\author{R.Simoniello}
\affil{Geneva Observatory, Chemin des Maillettes 51, CH-1290 Versoix, Switzerland}
\author{ S.~C.~Tripathy,  K.~Jain and F.~Hill}
\affil{National Solar Observatory, Tucson, AZ 85719, USA}
%\author{and et al.}
%\email{aastex-help@aas.org}
%\and
%\affil{CEA}


%\altaffiltext{1}{Visiting Astronomer, Cerro Tololo Inter-American Observatory.
%CTIO is operated by AURA, Inc.\ under contract to the National Science
%\altaffiltext{2}{Society of Fellows, Harvard University.}
%\altaffiltext{3}{present address: Center for Astrophysics,
%    60 Garden Street, Cambridge, MA 02138}
%\altaffiltext{4}{Visiting Programmer, Space Telescope Science Institute}
%\altaffiltext{5}{Patron, Alonso's Bar and Grill}


\begin{abstract}
The solar cycle onset at mid-latitudes, the slow down of the sunspot drift toward the equator, the tail-like attachment and the overlap of successive cycles at the time of activity minimum are delicate issues in $\alpha\Omega$ dynamo wave and flux transport dynamo models. Very different parameter values produce similar results, making it difficult to understand the origin of these solar cycle properties. We use GONG helioseismic data to investigate the progression of the solar cycle as observed in
intermediate-degree global $p$-mode frequency shifts at different latitudes and subsurface layers, from the beginning of solar cycle 23 up to the maximum of  the current solar cycle. We also analyze those for high-degree modes in each hemisphere obtained through the ring-diagram technique of local helioseismology.
The analysis highlighted differences in the progression of the cycle
below 15\degr\ compared to higher latitudes.
While the cycle starts at mid-latitudes and then
migrates equatorward/poleward, the sunspot eruptions of the old cycle are still ongoing below 15\degr\ latitude. This prolonged activity causes a delay in the cycle onset and an overlap of successive cycles, whose extension differs in the two hemispheres. Then the activity level rises faster reaching a maximum characterized by a single peak structure compared to the double peak at higher latitudes. Afterwards the descending phase shows up with a slower decay rate.
The latitudinal properties of the solar cycle progression highlighted in this study provide useful constraints to discern among the multitude of solar dynamo models.
 
\end{abstract}


\keywords{Sun: helioseismology --- Sun: dynamo}



\section{Introduction}
The cyclic behavior of solar magnetic activity is ascribed to the dynamo process powered by the inductive action of the turbulent fluid in the Sun's interior. A clear consensus has been reached on the $\Omega$ mechanism, which generates
toroidal field by shearing a pre-existing poloidal
field by differential rotation. Conversely it is still a matter of debate which $\alpha$ - effect 
regenerates poloidal fields from toroidal ones. There are two main competitive mechanisms: 1) the $\alpha$ turbulent effect, which
regenerates poloidal field from toroidal flux tube by helical motion \citep{Par55}; 2) the Babcock - Leighton mechanism, which is based on the observed decay of tilted, bipolar active regions, which acts as poloidal field sources at the surface \citep{Bab61, Lei64}. 
\textcolor{black}{While the $\alpha$ - turbulent $\Omega$ dynamo offered a plausible explanation for the sunspot drift toward equatorial latitudes by a dynamo wave, the Babcock - Leighton mechanism failed to reproduce the butterfly diagram. Therefore for several decades the $\alpha$-turbulent $\Omega$ dynamo has been favored over the Babcock - Leighton mechanism}. As various observations have found a poleward surface meridional flow \citep{Duv79,Kom93,Hat96}, the inclusion of a poleward circulation along with an equatorward subsurface return flow, 
initiated the development of so called Flux Transport Dynamo (FTD) model. This new class of dynamo models revived the Babcock - Leighton mechanism, as its inclusion in FTD   
models has been successful in reproducing many global solar cycle features \citep{Wan91,Dik99,Nan01}. {\textcolor{black}{The resulting simulations showed that the butterfly diagram is produced by the equatorward subsurface return flow advecting
the toroidal field toward the equator.
They have also been developed $\alpha$-turbulent $\Omega$ FTD models operating in the tachocline \citep{Dik01}\footnote{e.g. 
$\alpha$ turbulent effect is located at the base of the convection zone such as the models by \citet{Par93}
and \citet{Mac97}. These models are also known as thin shell or interface dynamo} and FTD simultaneously driven by an $\alpha$- turbulent effect located or in the tachocline or in the whole convection zone and a Babcock - Leighton type surface poloidal sources \citep{Bel13,Pas14}}. The key question is whether the Babcock - Leighton mechanism is an active component of the dynamo cycle, or a mere consequence of active region decay. How well the FTD and/or dynamo wave models reproduce the features observed in the butterfly diagram, might help solving the puzzle. 

 Fig.~\ref{fig:butt} shows the main features of the butterfly diagram: 1) the onset of cycle at mid-latitudes; 2) the sunspot drift toward the equator and its slow down
 represented by a change in the slope of the butterfly wing \citep{Mau04, Li01}; 3) the tail-like attachment over the minimum phase more prominent when the activity is stronger, which might lead to the overlap of successive cycles; \textcolor{black}{4) the length of the overlap varies within 1 - 2 years. It characterizes only the minimum phase and it is confined at latitudes $\le$ 15\degr\  \citep{Cli14}}.
 This feature is also seen in torsional oscillations shown in the bottom panels of Fig.~\ref{fig:butt} \citep{How09,Wil88}.
The sunspot drift rate toward the equator slows as the sunspot band approaches the
equator, and halts at about 8$\degr$ latitude \citep{Hat03}. The end of the migration does not correspond to the end of the activity as it produces the tail-like attachment. When the new cycle at mid-latitudes starts before the end of the old cycle at low latitudes, it causes the overlap of successive cycles. \textcolor{black}{FTD models driven only by Babcock - Leighton mechanism \citep{Cha04} or along with the $\alpha$ -turbulent effect operating in the bulk of the convection zone, currently, have the best agreement with observations \citep{Pas14}, as the length of the simulated overlap is short and it occurs only during the minimum at low latitudes. Conversely, thin shell dynamo wave models \citep{Mos00,Bus06,Sch04} or flux transport thin shell dynamo \citep{Dik01}, tend to produce dynamo waves with too short wavelength leading to excessive overlap between
adjacent cycles as it involves a wider range of latitudes. Furthermore they also fail to reproduce the tail - like attachment over the minimum phase.} 
Moreover the direction of the activity migration could also provide information on the nature of the $\alpha$ mechanism. Both formalisms make strong assumptions to initiate the sunspot cycle at mid-latitudes. The Babcock - Leighton FTD models assume that the deep equatorward meridional flow penetrates slightly below the convection zone to a greater depth than usually believed \citep{Nan02}, to prevent the onset and occurrence of  a sunspot  cycle above 45$\degr$ as well as any other kind of cyclic activity. %; \citep{Nan02}. 
The same result is achieved in $\alpha\Omega$ dynamo wave by inhibiting the $\alpha$ - turbulent effect at higher latitudes \citep{Sch04}. %The absence of any cyclic magnetic activity above 45$\degr$ does not match helioseismic observations \citep{Simo13}. 
Based on these assumptions in any type of FTD models the magnetic activity starts at higher latitudes and then propagates only equatorward, while in thin shell $\alpha\Omega$ dynamo wave the magnetic activity can propagate equatorward as well as poleward \citep{Bus06}. {These two branches are also clearly seen in
the torsional oscillation pattern \citep[e.g.][]{How09}. It results from the solar - like differential profile, which is characterized by a sign change in $\frac{\delta\Omega}{\delta r}$ at high - and low latitude tachocline \citep{Rud95}. This sign change, however has not yet been confirmed by helioseismic observations. 

In this work we aim at characterizing
%we address the role of the Babcock - Leighton mechanism with respect to the $\alpha$ turbulent effect by characterizing 
the different phases of solar cycle at all latitudes and in the two hemispheres, as these properties can be used to constrain solar dynamo models. }%and comparing our findings with the predictions of the $\alpha\Omega$ dynamo wave and FTD models. 
We use acoustic $p$-mode frequencies as a diagnostic tool to infer the progression of the 11 yr magnetic cycle. They are very well known to 
 correlate strongly with solar magnetic activity {\citep{Els90, Lib90, How99, Jai01, Sim10}} and unlike to many other solar activity proxies they probe magnetic changes induced by weak as well as strong toroidal fields at all latitudes. In order to simultaneously track solar magnetic activity in both hemispheres separately, we further use localized high-degree frequencies
from the ring-diagram technique. The paper is organized as follows: in Sect. 2 we describe the data analysis for both intermediate and high-degree mode. The results are presented in Sect. 3 followed by the comparison with sunspot numbers in Sect. 4, and the findings are discussed in Sect. 5.
 \begin{figure*}
\centering
\includegraphics[scale=.2]{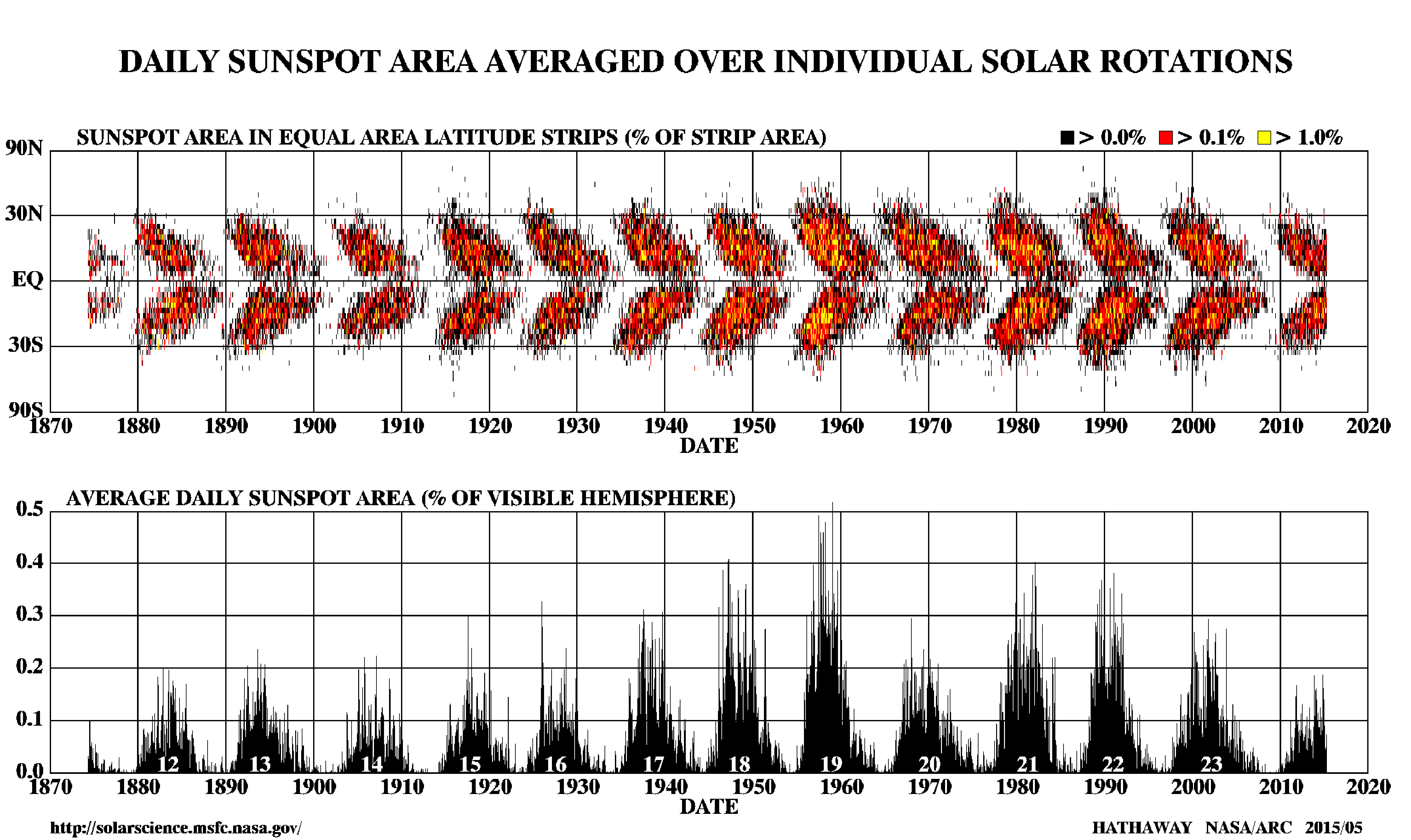}
\includegraphics[scale=.4, trim=0 0 0 0]{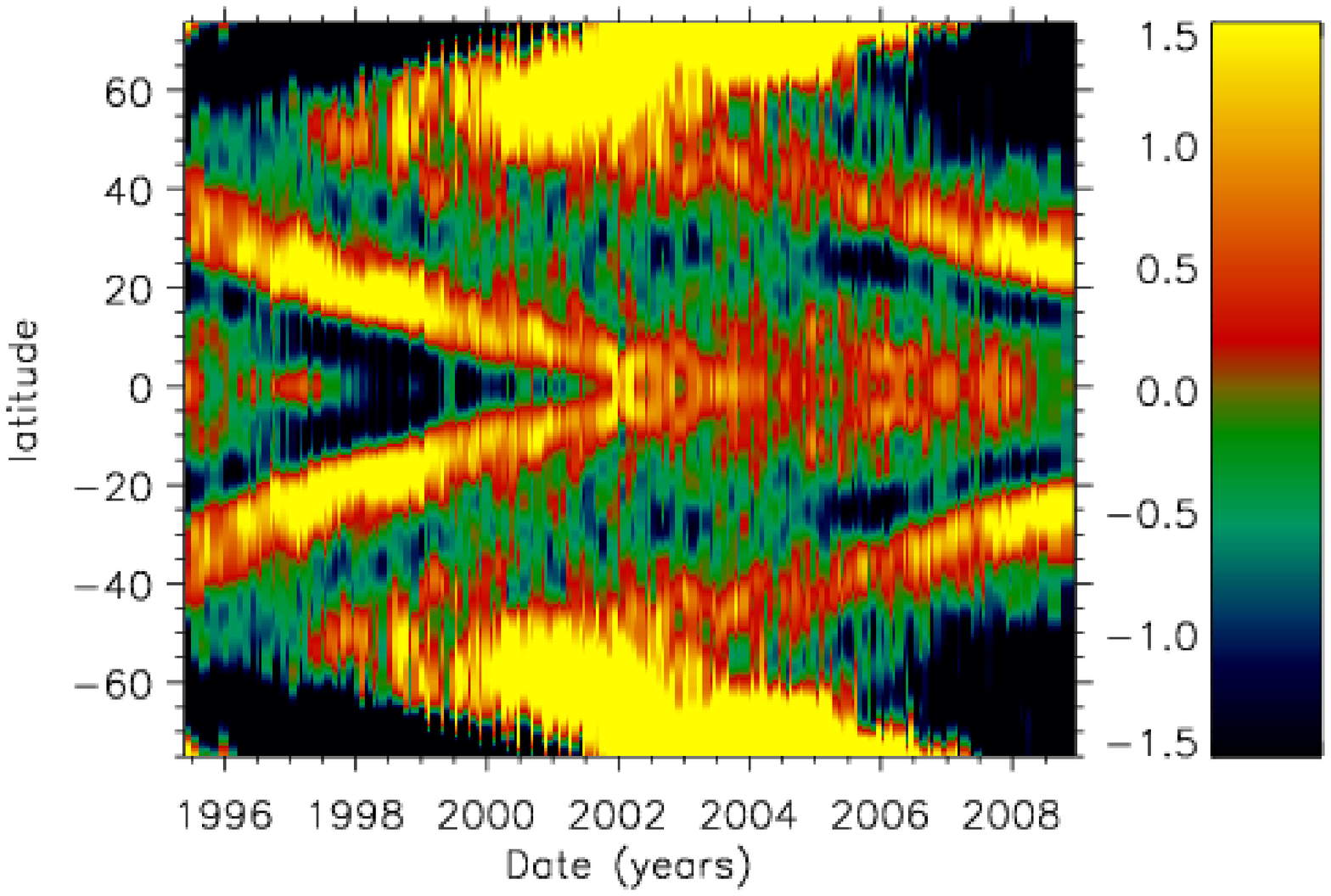}
\includegraphics[scale=.4, trim=0 0 0 0]{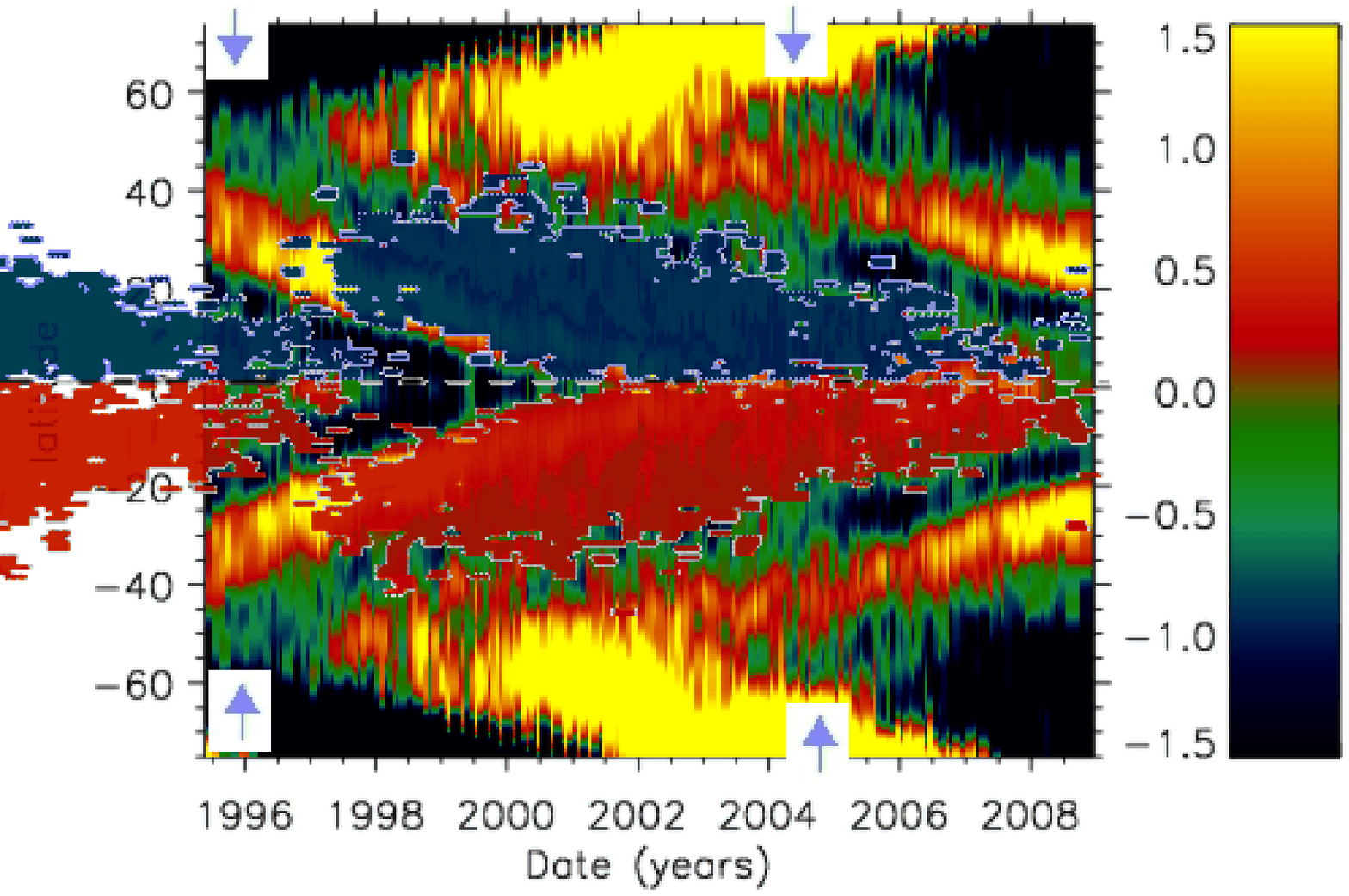}
\caption{Top two panels show the butterfly diagram and the Sunspot area daily averaged (taken from www.msfc.nasa.gov).
Bottom left panel show the torsional oscillations from \cite{How09} and bottom right the same with the over plotted butterfly diagram provided by Solar Influences Data Center (SIDC) using sunspot numbers and rescaled to match the left panel. { The plot extends outside the axes as the SIDC sunspot data have been over plotted since the descending phase of solar cycle 22, when GONG observations did not yet started (taken from https://landscheidt.wordpress.com/2009/02/25/latest-solar-differential-rotation-information/)}.The arrows identify the begin of the equatorward branch at around 40\degr\ .}
\label{fig:butt}
\end{figure*}

\section{Data Analysis}
\subsection{The GONG data}
 In this work we look for temporal variations in $p$-mode frequencies caused by changes in magnetic 
activity levels as function of latitude and subsurface layers. The mode frequencies analyzed here are obtained from 
the Global Oscillation Network Group (GONG) \footnote{ftp://gong.nso.edu/data/} in two different degree ranges.
The low-and intermediate-degree global mode frequencies are obtained 
  for the individual ($n, \ell, m$) 
multiplets, $\nu_{n \ell m}$, where $n$ is the radial order and $m$ is the
azimuthal order, running  from $-\ell$ to $+\ell$.  The mode frequencies for each 
multiplet were estimated from the $m-\nu$ power spectra constructed by the time 
series of 108 days. The data analyzed here consist of 
overlapping data sets, with a spacing 
of 36 days between consecutive time series,  covering the
period from June 1995 to July 2013
 in the $\ell$ range of 20 $\le \ell \le$ 147 and frequency range 1500$\mu$Hz $\le\nu\le$ 3900$\mu$Hz. 
 
\textcolor{black}{The high-degree mode frequencies  are obtained from the localized regions (15\degr $\times$ 15\degr) 
on the solar surface using the GONG ring-diagram pipeline \citep{sct-corbard03}. The analysis covers a period from 2001 
July to 2014 June in the degree range of  180 $\le \ell \le$ 1000.  In the ring-diagram method,  the localized regions 
on the solar surface are tracked with an average rotation rate at the solar surface for 1664 minutes. 
 Each tracked area is apodized with a circular function and then a three-dimensional FFT is applied on 
both spatial and temporal direction to obtain a three-dimensional power spectrum. Finally, the corresponding power spectrum 
is fitted using a Lorentzian profile model to obtain acoustic mode parameters. The high-degree modes provide information
about the outermost layer of the Sun's interior.}

\subsection{Determination of the frequency shifts }
We aim at characterizing the progression of solar cycle at different latitudes. The sunspot cycle starts at mid-latitudes ($\approx 30\degr$ latitude), it reaches the maximum at $\approx$ 15\degr\ latitude and stops at around 8\degr\ latitude. \textcolor{black}{Modes of low to intermediate degree $\ell$ are global, and sense the spherical geometry of the Sun. Therefore these are better described by spherical harmonics of the form $Y_{\ell, m}(\theta,\phi)= P_{\ell,m}(cos\theta)e^{i m\phi}$, where P is the Legendre Polynomial, $\ell$ the spherical degree and $m$ the azimuthal order. The spherical harmonic degree $\ell$ is the number of nodes along a circle at an angle $\theta=arccos\frac{m}{\sqrt(\ell(\ell+1)}$ at the equator. The azimuthal order $m$ is the number of nodal lines crossing the equator.  
We, therefore, used the above ratio to select acoustic modes depending on their upper latitude range to which they are more sensitive. It may be noted that acoustic modes with same spherical degree $\ell$, but different azimuthal order $m$ increase their sensitivity to lower latitudes with increasing $m$. In fact
the sectoral modes ($|{m}|$ = $\ell$) are more sensitive to the regions near the equator while the zonal modes ($m$ = 0) have higher sensitivity at higher 
latitudes \citep{hill91}.} We carry on this selection in five latitude ranges between 0\degr\ $\le\theta\le$ 75\degr\ spaced by 15\degr\ .
This allow us to split the cycle progression in latitude ranges. {Since the acoustic waves travel throughout the interior and they reflect back from different layers depending on
their frequencies, we further investigate the progression of the solar cycle based on their upper turning point ($u_{p}$). 
With increasing frequency $u_{p}$ approaches the surface \citep{Bas12}. We divide frequency data sets
in to three groups: 
(i) low-frequency range 1500~$\mu$Hz $\le\nu\le$ 2300~$\mu$Hz correspondes to 0.9944~R$_{\odot}$ $\le u_{p}\le$ 0.9987~R$_{\odot}$,
(ii) medium-frequency range 2300~$\mu$Hz $\le\nu\le$ 3100~$\mu$Hz for 0.9987~R$_{\odot}$ $\le u_{p}\le$ 0.9998 R$_{\odot}$,
and (iii) high-frequency range 3100~$\mu$Hz $\le\nu\le$ 3900~$\mu$Hz for  0.9998~R$_{\odot}$ $\le u_{p}\le$  0.9999~R$_{\odot}$.
% Furthermore with increasing frequency the upper turning point ($u_{p}$) approaches the surface as follows: 1) 0.9944~R$_{\odot}$ $\le u_{p}\le$ 0.9987~R$_{\odot}$ in the low frequency range (LFR) between 1500~$\mu$Hz $\le\nu\le$ 2300~$\mu$Hz; 2) 0.9987~R$_{\odot}$ $\le u_{p}\le$ 0.9998 R$_{\odot}$  in the medium frequency range (MFR) between
%2300~$\mu$Hz $\le\nu\le$ 3100~$\mu$Hz; 3) 0.9998~R$_{\odot}$ $\le u_{p}\le$  0.9999~R$_{\odot}$ in the high frequency range (HFR) between 3100~$\mu$Hz $\le\nu\le$ 3900~$\mu$Hz \citep{Bas12}. 
As helioseismic observations have shown that the size of the frequency variation with the solar cycle 
 increases as $u_{p}$ approaches the surface \citep{Cha01, Simo13},
 we then calculated the frequency shifts in the low, medium and high frequency band, to investigate the solar cycle properties in different subsurface layers. %Helioseismic observations, in fact, have shown that the size of the shifts increases as we approach the surface \citep{Cha01,Simo13}.
Mode frequency shifts $\delta\nu_{n,\ell}(t)$ were
defined as  the differences between the frequencies observed at
different times ($\nu_{n,\ell}(t)$) and the reference values of the corresponding modes ($\nu(ref)$):
\begin{equation}
\delta\nu_{n,\ell}(t)=\nu_{n,\ell}(t)-\nu_{n,\ell}(ref)
\end{equation}
%  the differences between the frequencies observed at
%different times ($\nu_{n,\ell}(t)$) and the reference values of the corresponding modes ($\nu_{ref}$).
The $\nu_{n,\ell}(ref)$ was determined 
 as the average frequency 
 over the minimum between cycle 22 and 23. We took into account the period of observations from June 1995 up to May 1996.  { While we have included the end of activity cycle 22, we have been very careful not to include in $\nu_{n,\ell}(ref)$ the beginning of solar activity at higher latitudes, as it would have led significant differences in the size of $\nu_{n,\ell}(ref)$ at different latitudes making difficult any comparison and interpretation of the solar cycle properties}. We then determine the weighted frequency difference in the low, medium and high frequency band for each selected latitude. The weights ($\frac{1}{\sigma^{2}}$), are the errors of the fitting procedure.}

The high-degree mode frequency shifts were obtained by analyzing the frequencies obtained from from ring-diagram technique.
These frequencies in localized regions are affected by the foreshortening as well as the gaps in observation, thus
 have been corrected by modeling these effects as a two-dimensional function of the distance from the disk center and a linear dependence of the duty cycle
\citep{howe-apj04,sct-sol13}. The corrected frequencies are then used to compute frequency shifts  which is computationally similar to those of 
global modes except for the choice of the reference frequency. For high-degree modes, the frequency difference of each mode is computed with respect to the average frequency of the same mode over the 189-dense pack tiles (covering $\pm$ 60\degr on the disk) corresponding to a magnetically quiet day (2008 May 11). In a similar manner  frequency shifts for each hemisphere 
and at different latitude ranges were computed with an appropriate reference frequency as described in \citet{sct-apj15} e.g. for northern hemisphere, the reference frequency was computed only over the northern hemisphere.
 \section{Results}
\subsection{Progression of solar cycle at different latitudes}
% Since the frequency shifts are strongly correlated with magnetic activity indices over the time scales of solar cycle, it has been argued by \cite{How99} that the shifts may be considered as a proxy of solar magnetic activity. %We characterize the shifts in terms of measure of the activity in rest of the paper.
 {Since numerous examples have clearly shown the strong correlation between  frequency shifts and the magnetic activity indices over
 different time scales, %we  has been argued that the shifts may be considered as a proxy of solar magnetic activity. 
 we interpret frequency shifts as a measure of solar activity in rest of the paper.}

 \begin{figure*}
 \centerline{
 \includegraphics[scale=.5]{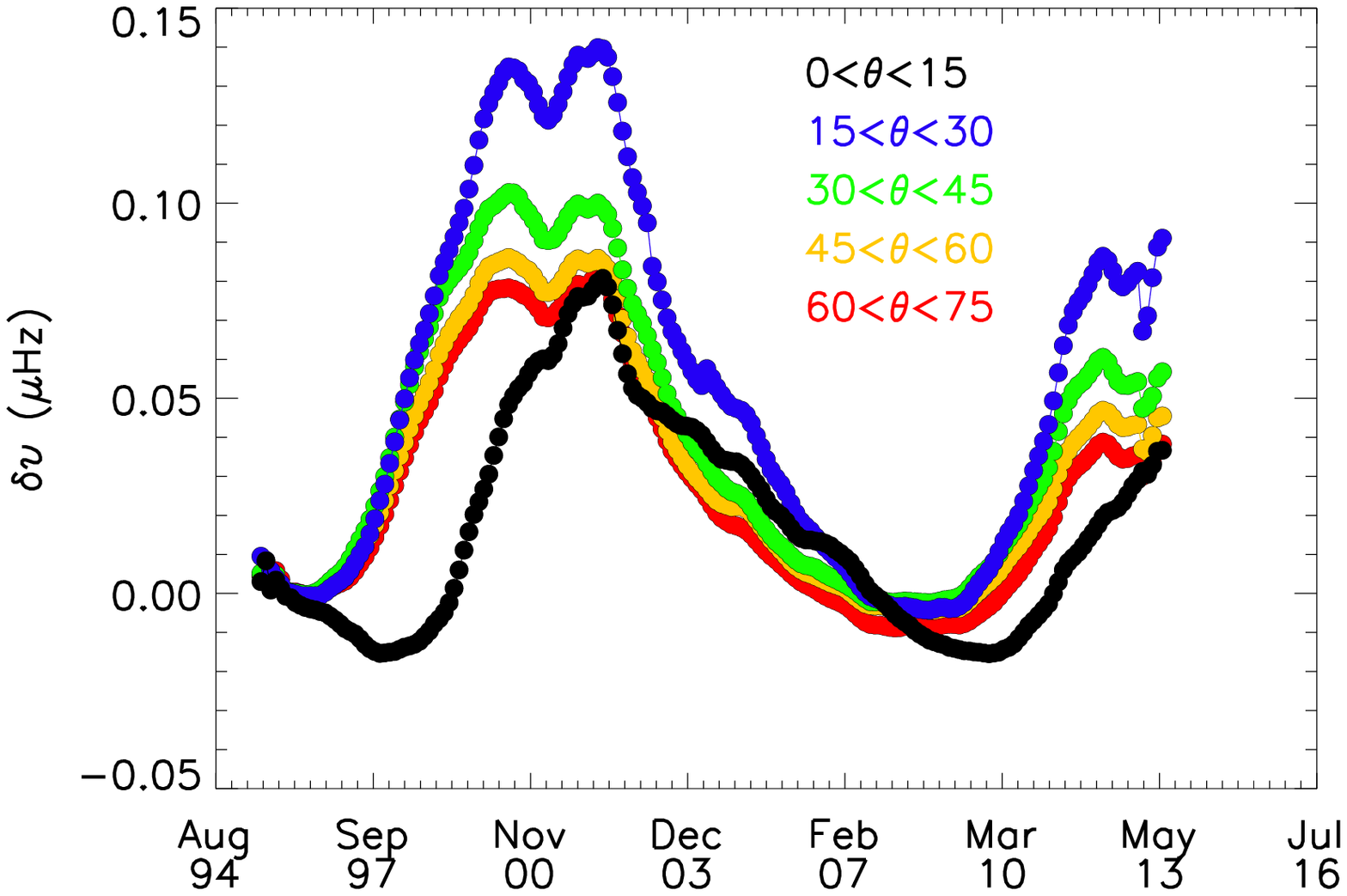}
 }
 \centerline{
\includegraphics[scale=.45]{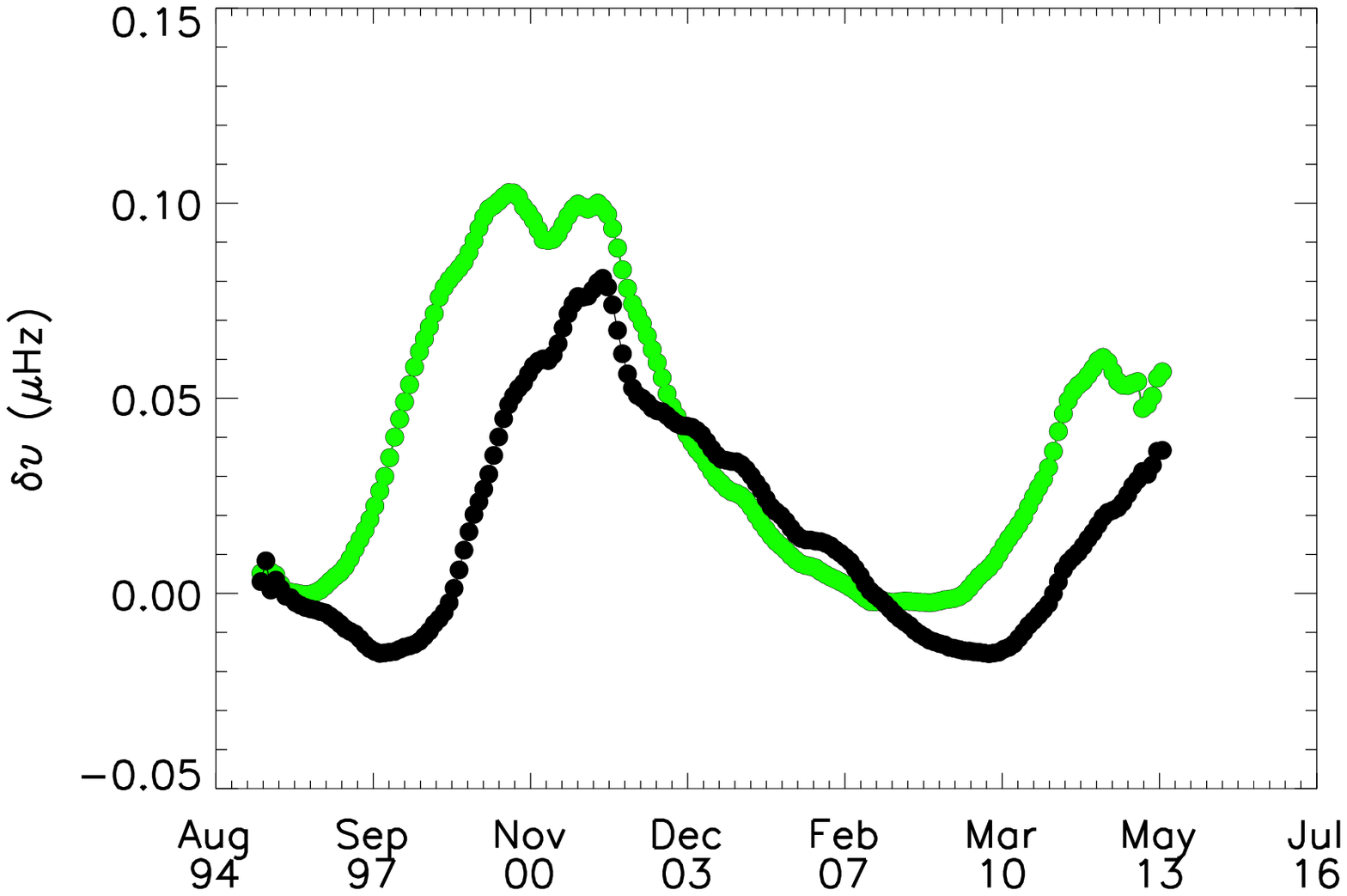}
\includegraphics[scale=.45]{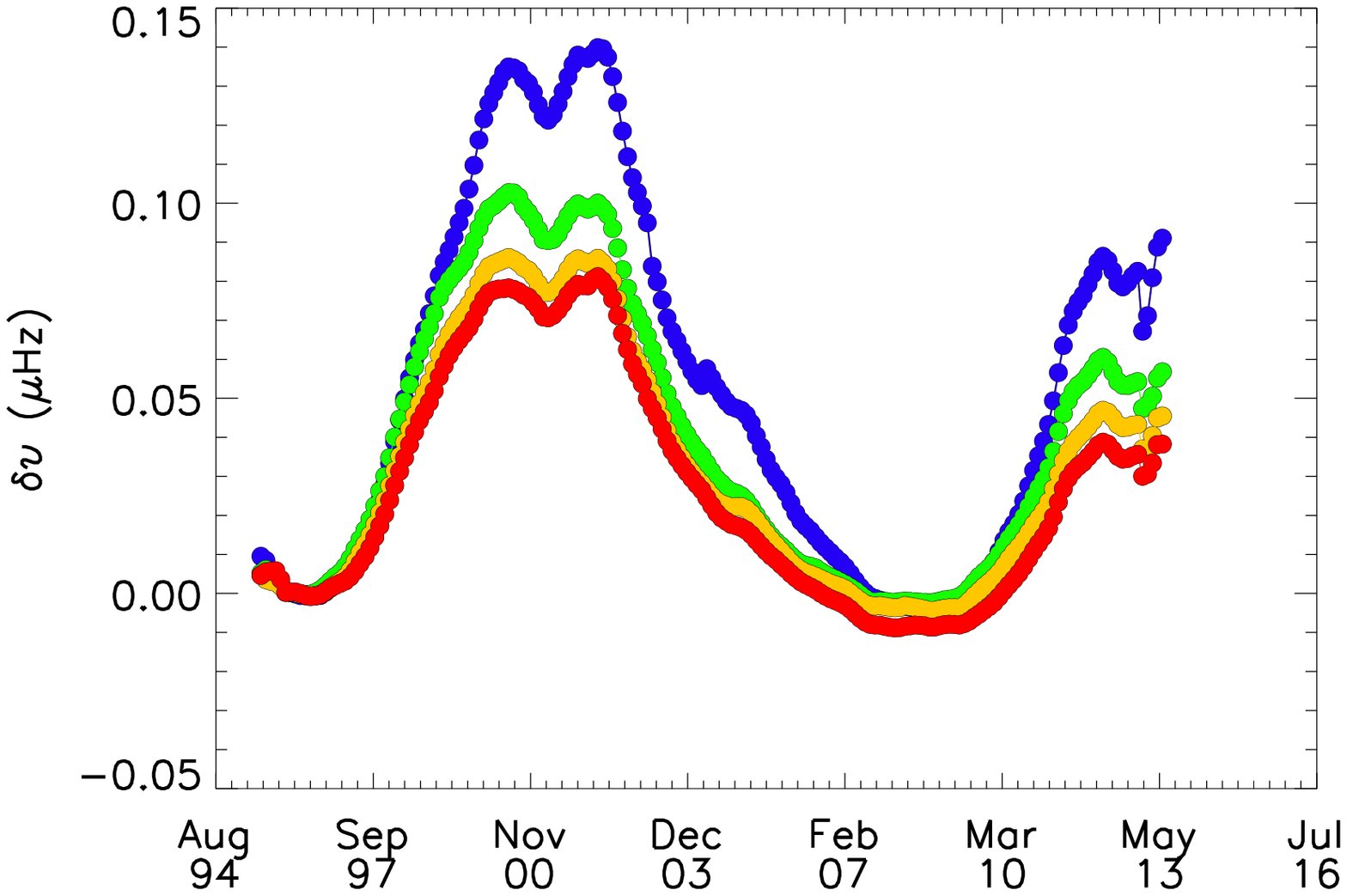}}
\caption{Top panel shows the variation of $p$-mode frequency shifts with time at five different latitudinal bands. 
Bottom row: left panel highlights the differences in the evolution of magnetic activity as measured by the
frequency shifts during the descending phase between 0\degr $\le\theta\le$ 15\degr\ (black) and 
30\degr $\le\theta\le$ 45\degr\ (green)  while right panel shows the similarities in four latitude bands.}
\label{fig:fig_allfreq_0_15}
\end{figure*}

{Top panel of Figure~\ref{fig:fig_allfreq_0_15} shows the variation of frequency shifts over solar cycle 23 and the ascending phase of solar cycle 24 at 
five latitude bands in the frequency range of 1500~$\mu$Hz $\le\nu\le$ 3900~$\mu$Hz. All curves have been smoothed with a boxcar of 1 year and 
the estimated uncertainties  are of the order 10$^{-3}\mu$Hz. It is clearly seen that the progression of activity cycle is different at 
different latitudes.  In particular, the variation of the magnetic activity below 15\degr\ differs from the one at higher latitudes; 
it rises faster, the maximum is characterized by a single peak structure and  an excess of activity changes the evolutionary path 
of the descending phase around 2003 December.  This difference is better highlighted in the bottom left panel of Figure~\ref{fig:fig_allfreq_0_15},
which compares the activity at 0\degr\ $\le\theta\le$ 15\degr\ with 30\degr\ $\le\theta\le$ 45\degr.  At 0\degr\ $\le\theta\le$ 15\degr, the declining phase lasted longer 
compared to higher latitudes delaying the time of the minimum and the onset of solar cycle 24 for about a year, which led to an 
overlap of cycle 23 with cycle 24. Similar longer delay in the onset of solar cycle 23 caused an overlap between cycle 22 and 23.
Furthermore over both minimum phases, the activity level reached its deepest value at latitudes between 0\degr $\le\theta\le$ 15\degr. 
The bottom right panel highlights the 
similarity in the progression of solar activity at all latitudes above 15\degr. Here, the activity level at both minimum phases 
at all latitude bands is of comparable size. It rises with slightly different growth rates and
reaches the maximum characterized by the typical double peak structure, which has been interpreted as a manifestation of the Quasi-Biennial
Periodicity \citep[QBP;][]{Fle10, Jai11,Sim12a,Simo13}. Soon after the second maximum, the descending phase continued with comparable decay times, 
although around 2003 December  the progression at latitudes between 15\degr $\le\theta\le$ 30\degr\ slightly changed.

To summarize similarities and differences in the properties of the progression of  solar cycle at different latitudes and in the low, medium and high frequency range, 
Table 1 lists (i) the epochs of minimum and maximum of solar cycle in each latitudinal 
band; these have been defined as the timing corresponding to the lowest/highest value in the frequency shift at different latitudes, (ii) the rising and decay time, (iii) the full cycle length.
The solar  cycle progression shows common features in the three bands:
starts within the latitude range 30\degr\ $\le\theta\le$ 45\degr\ and followed by other latitudes. Between 0\degr\ $\le\theta\le$ 15\degr\ latitude the onset of the 
new cycle is always delayed,  %but we note a difference by more than one year in the medium and high frequency range, and by almost two years in the low frequency range, 
but this time lag disappears at the time of the maximum, as it occurs at the same epoch at all latitudes in each frequency range (Max(2) in Table 1).
%This implies that the ascending phase below 15\degr\ latitude is always faster compared to higher latitudes. This feature characterizes all the three frequency ranges.
This result further confirms that the rise in activity below 15\degr\ latitude is faster compared to all other latitudes (Rise time (2) in Table 1). %This feature characterizes all the three frequency ranges.
  The descending phase, instead, lasted longer below 15\degr\ latitude compared to higher ones, leading to a slow down of the progression of magnetic activity, which ended in the overlap of successive cycles (Decay time in Table 1).  This peculiar behavior ended up in a stronger asymmetry between the rise and decay time at latitudes below 15\degr\ compared to higher ones.
%It is also important to highlight one further point: in the low frequency band there is no difference in the decay time between 15\degr\ $\le\theta\le$ 30\degr\  and 30\degr\ $\le\theta\le$ 45\degr\. This does not occur in the medium and him frequency range, where instead we note a difference of almost one year in the decay time.
Even though the descending phase lasted longer at latitudes $\le$ 15\degr\ , the fastest rising phase made cycle lengths of comparable size at all latitudes in all frequency bands (Length in Table 1). 
%The last point to highlight is the onset of cycle 24 at latitudes between 15\degr\ $\le\theta\le$ 30\degr\ and in the medium and high frequency range. We note that compared to higher latitudes the cycle onset has been delayed by almost one year (Cycle 24/Min Table 1). This feature does not occur in the onset of cycle 23.% delayed compared to higher latitudes in compar%Differences in the decay time and cycle lengths of almost one year are seen at latitudes above 15\degr\ between the low frequency and the medium and high frequency range.
%The last point to highlight is that in all frequency bands the rise and decay time at all latitudes above 15\degr\ , %although slightly quicker is the decay time at latitudes  30\degr\ $\le\theta\le$ 45\degr\ .% significant differences, .  
%These findings further support the existence of two distinct modes of activity, which converge at the time of the maximum.} %which might be driven or by the same mechanism or by two different mechanisms.%, a  characterized by a fast growing and slow decay, or by a different process. % that during the ascending and descending phase  the activity at low latitudes compared to higher ones.}
\begin{table*}
\begin{center}
%\small\addtolength{\tabcolsep}{-3pt}
\caption{\textcolor{black}{The solar cycle onset, peak amplitude, rise and decay time, and the full cycle length in the low, medium and high frequency range.}}
%\small\addtolength{\tabcolsep}{-3pt}
\scalebox{1.2}{
\begin{tabular}{|c|c|c|c|c|c|c|c|c|}
\tableline\tableline
\multicolumn{9}{|c|}{\textcolor{red}{Low Frequency Range}}\\ \hline
Lat&\multicolumn{3}{c|}{Cycle 23}&Cycle 24& Rise (1) & Rise (2) &Decay &Length \\ \hline
Degree&Min&Max (1)& Max (2)&Min&Months&Months& Max-Min&Months\\ \hline
%\multicolumn{9}{|c|}{LFR}\\ \hline
0$\le\theta\le$15& 02/1998&                &03/2002 &03/2010&    &49 & 96 &145\\ \hline
15$\le\theta\le$30&03/1996&06/2000&03/2002&11/2007& 51&72 & 68 &140\\ \hline
30$\le\theta\le$45&01/1996&05/2000& 03/2002&07/2007&52&74 & 64&138\\ \hline
45$\le\theta\le$60&08/1996& 04/2000& 03/2002&07/2007&44&67 & 64&131 \\ \hline
60$\le\theta\le$75&09/1996& 04/2000&03/2002&08/2007&43&66& 65&132\\ \hline
\multicolumn{9}{|c|}{\textcolor{red}{Medium Frequency Range}}\\ \hline
0$\le\theta\le$15&11/1997&                 &04/2002 &01/2010&    &53 &93 &146\\ \hline
15$\le\theta\le$30&09/1996&07/2000&04/2002&04/2009& 46&  67 &84 &151\\ \hline
30$\le\theta\le$45&06/1996&06/2000 &03/2002&05/2008& 48 & 69  &74&143\\ \hline
45$\le\theta\le$60&07/1996&06/2000&03/2002&11/2008& 47 & 68 &80&148 \\ \hline
60$\le\theta\le$75&08/1996&05/2000&03/2002&12/2008& 45 &67 &81&148\\ \hline
\multicolumn{9}{|c|}{\textcolor{red}{High Frequency Range}}\\ \hline
0$\le\theta\le$15&12/1997&                &04/2002 &01/2010&      &52 &93 &145\\ \hline
15$\le\theta\le$30&09/1996&06/2000&03/2002&03/2009& 45& 66    &84 &150\\ \hline
30$\le\theta\le$45&09/1996&06/2000&03/2002&10/2008& 45 &  66  &79&145\\ \hline
45$\le\theta\le$60&10/1996&06/2000&03/2002&11/2008& 44 &  65  &80&145 \\ \hline
60$\le\theta\le$75&11/1996&06/2000&03/2002&12/2008& 43 & 64 &81&145\\ \hline
%75$\le\theta\le$90&07/1996&03/2002&03/2009&12.7\\ \hline
\tableline
\end{tabular}}
\label{tab:time_minimum}
\end{center}
\end{table*}
\subsection{Sensitivity of the progression of solar cycle to the subsurface layers}
To further study whether such behavior persists in the solar subsurface layers, each panel of Fig.~\ref{fig:fig_allfreq} 
 shows the progression of solar cycle in a selected frequency band and at all latitudes. %The lower $\frac{nu}{\ell}$ range
%corresponds to modes confined to the deeper layers while the higher $\frac{\nu}{\ell}$ is for modes that penetrate shallower layers.
 As we can see the properties of the two different modes of the solar cycle as described above do not change. %Over the minimum again at all depths the activity is lowest below 15$\degr$.  
% \textcolor{black}{Low and intermediate degree mode analysis has highlighted that, during solar cycle 23, the activity was further characterized by the 2 yr signal (QBP) \citep{Sal09, Fle10, Simo13}. In particular the QBP signal seemed to have better characterized the solar activity during descending phase rather than the ascending one, around the years  2003 - 2007. Comparing the progression of activity in the three subsurface layers over the period December 2003 and February 2007, we note that, as we go deeper in the Sun, a further modulation of almost 2 years seems to characterize the descending phase of cycle 23. It has been shown that the size of the shift over the 11 and 2 yr cycle decreases with decreasing frequency, and differences  in the size of the shifts over the two cycles tend to become smaller as we go deeper \citep{Simo13}.}
%What instead seems to become more prominent over the descending phase is the amplitude modulation of about 2 years linked to the QBP signal, as the size of the shift over the 11 and 2 yr cycle becomes of comparable  strength \citep{Simo13}.
\begin{figure*}
\centerline{
\includegraphics[scale=.48]{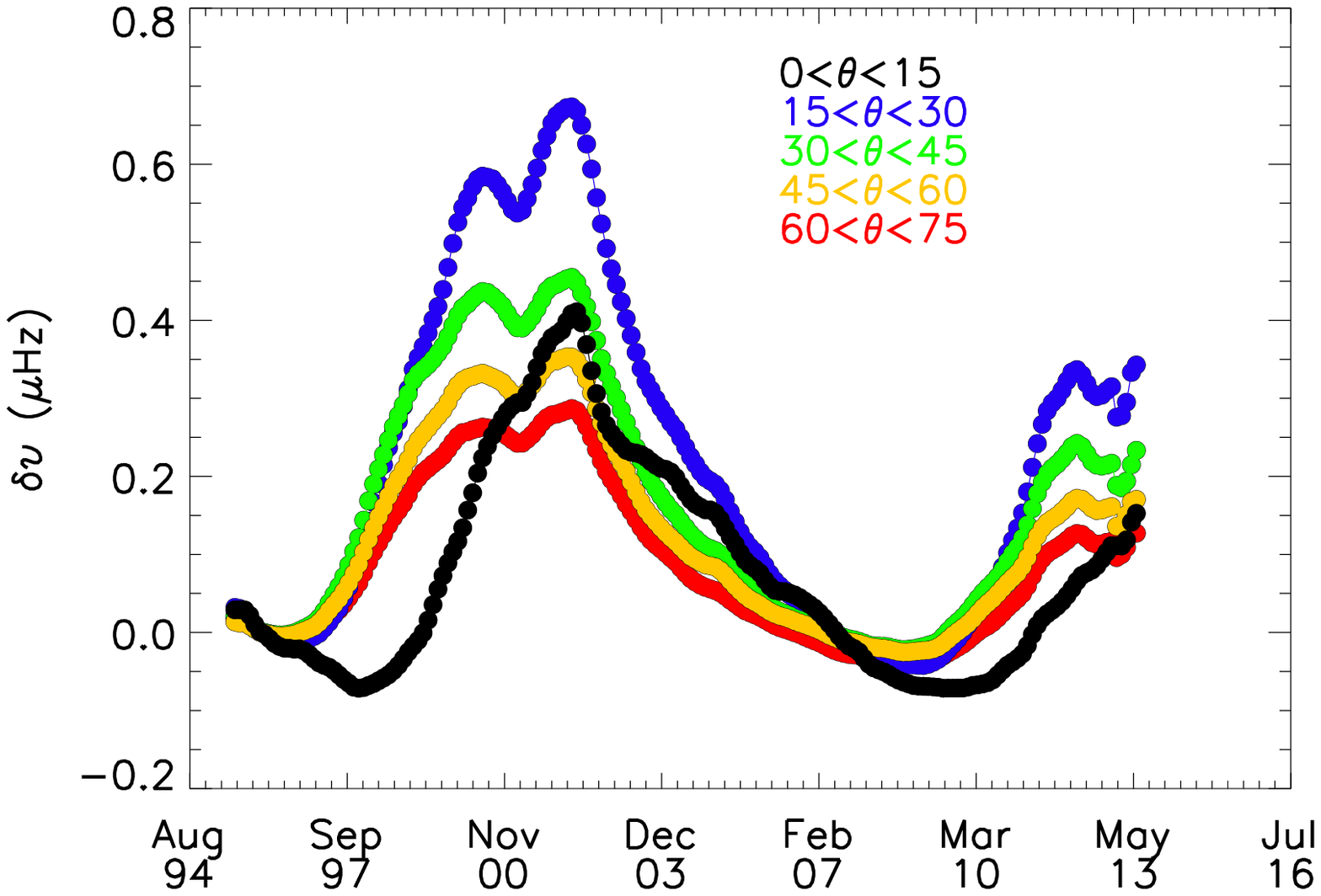}
}
\centerline{
\includegraphics[scale=.48]{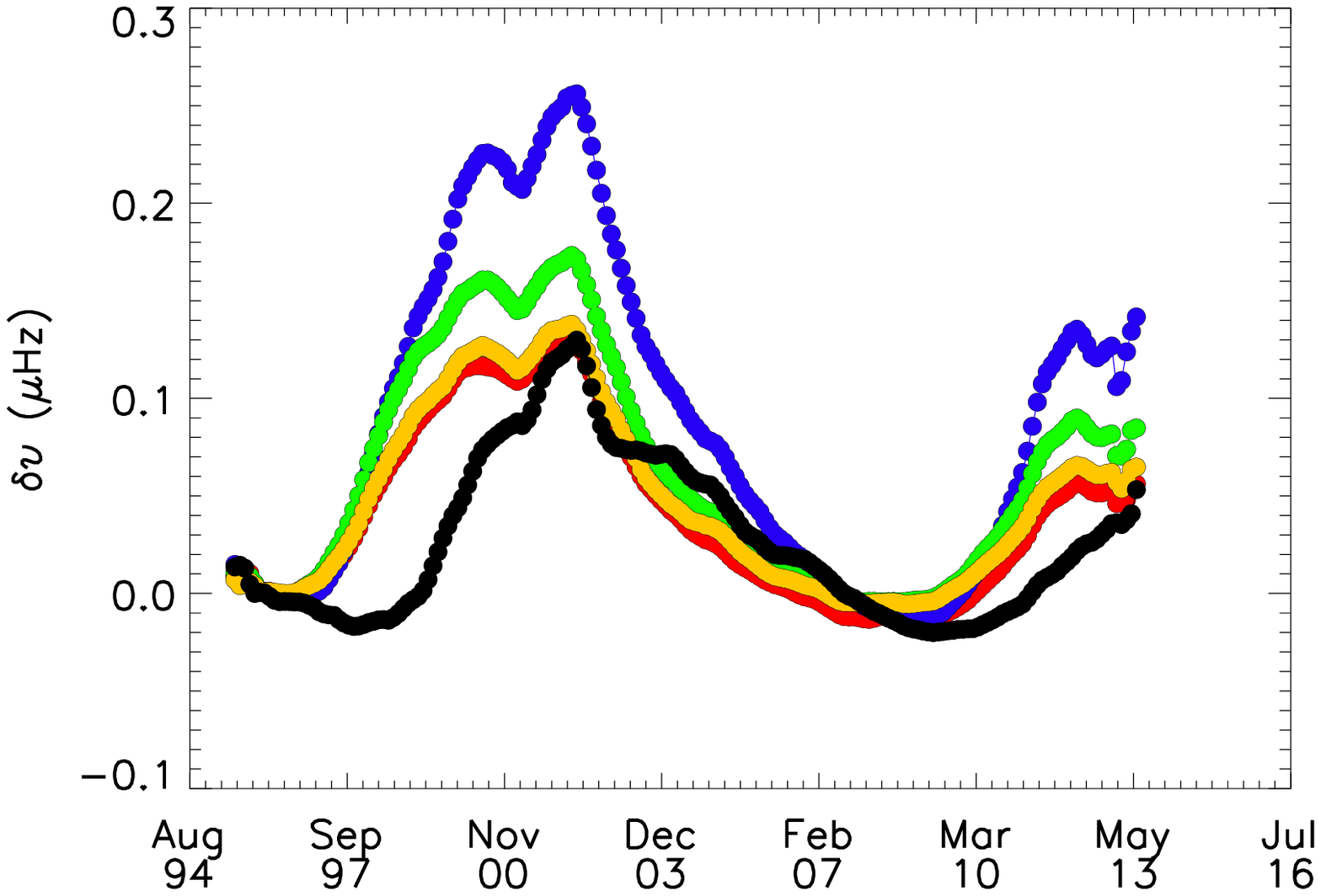}}
\centerline{
\includegraphics[scale=.48]{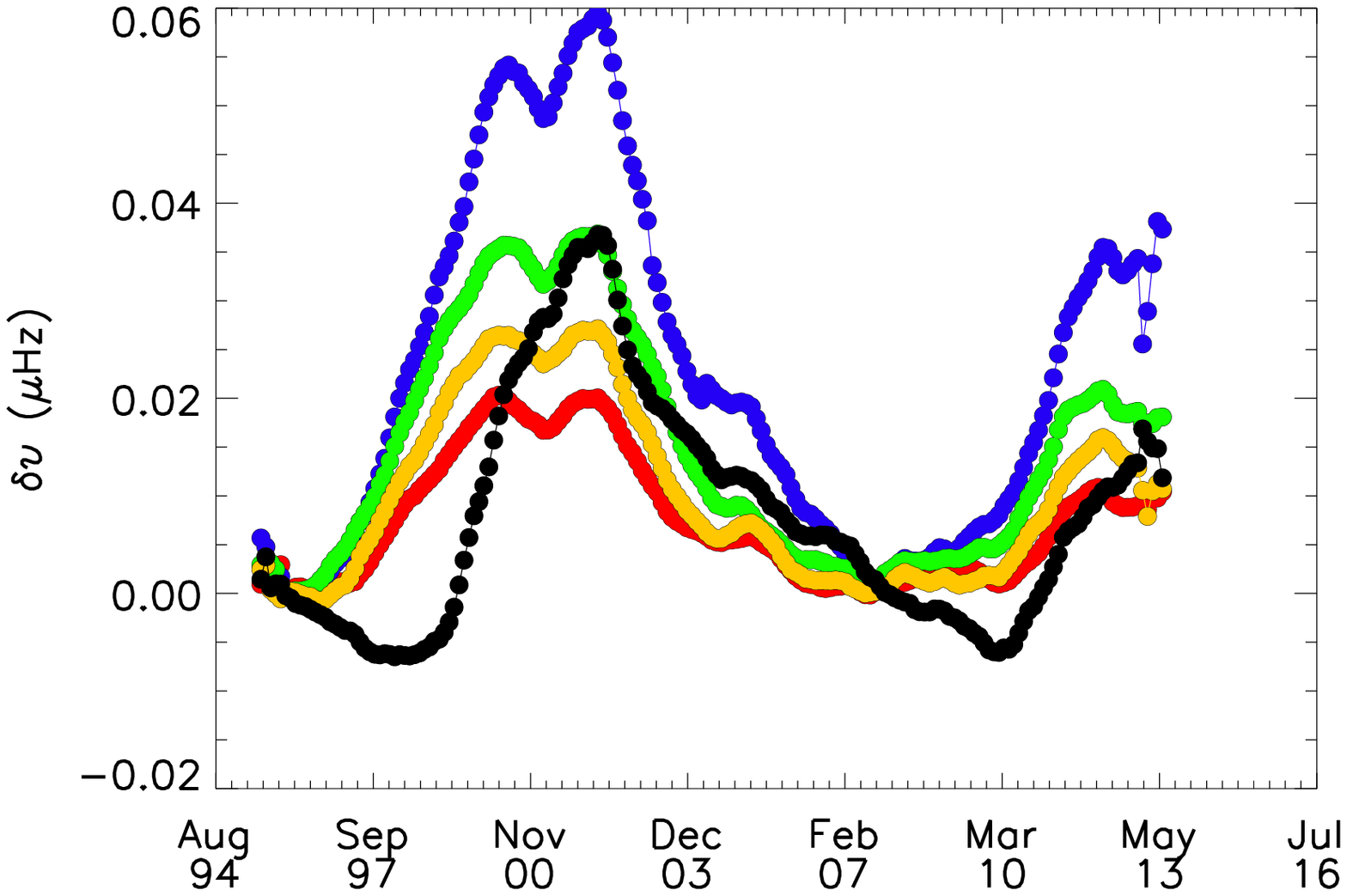}}
\caption{Solar cycle changes in $p$-mode frequency shifts at five different latitudinal bands in three different 
frequency ranges: (top) 3100~$\mu$Hz $\le\nu\le$ 3900~$\mu$Hz, (middle)  2300~$\mu$Hz $\le\nu\le$ 3100~$\mu$Hz,
and (bottom)  1500~$\mu$Hz $\le\nu\le$ 2300~$\mu$Hz. Note that the scales on y-axis in all three plots are different. }
\label{fig:fig_allfreq}
\end{figure*}

{\textcolor{black}{  To better compare the behavior and strength of activity in the three subsurface layers, each panel of Fig.~\ref{fig:difffreq_0_45} compares the activity at the same latitude but in the three frequency ranges. As expected the activity is stronger in the nearest subsurface layers compared to deeper ones.} %Over the minimum phase, instead, at latitudes above 15\degr\ the activity is almost comparable in the three frequency ranges, while a weaker activity level can be seen below 15\degr\  in the higher frequency band.} 
\begin{figure*}
\centerline{
\includegraphics[scale=.45]{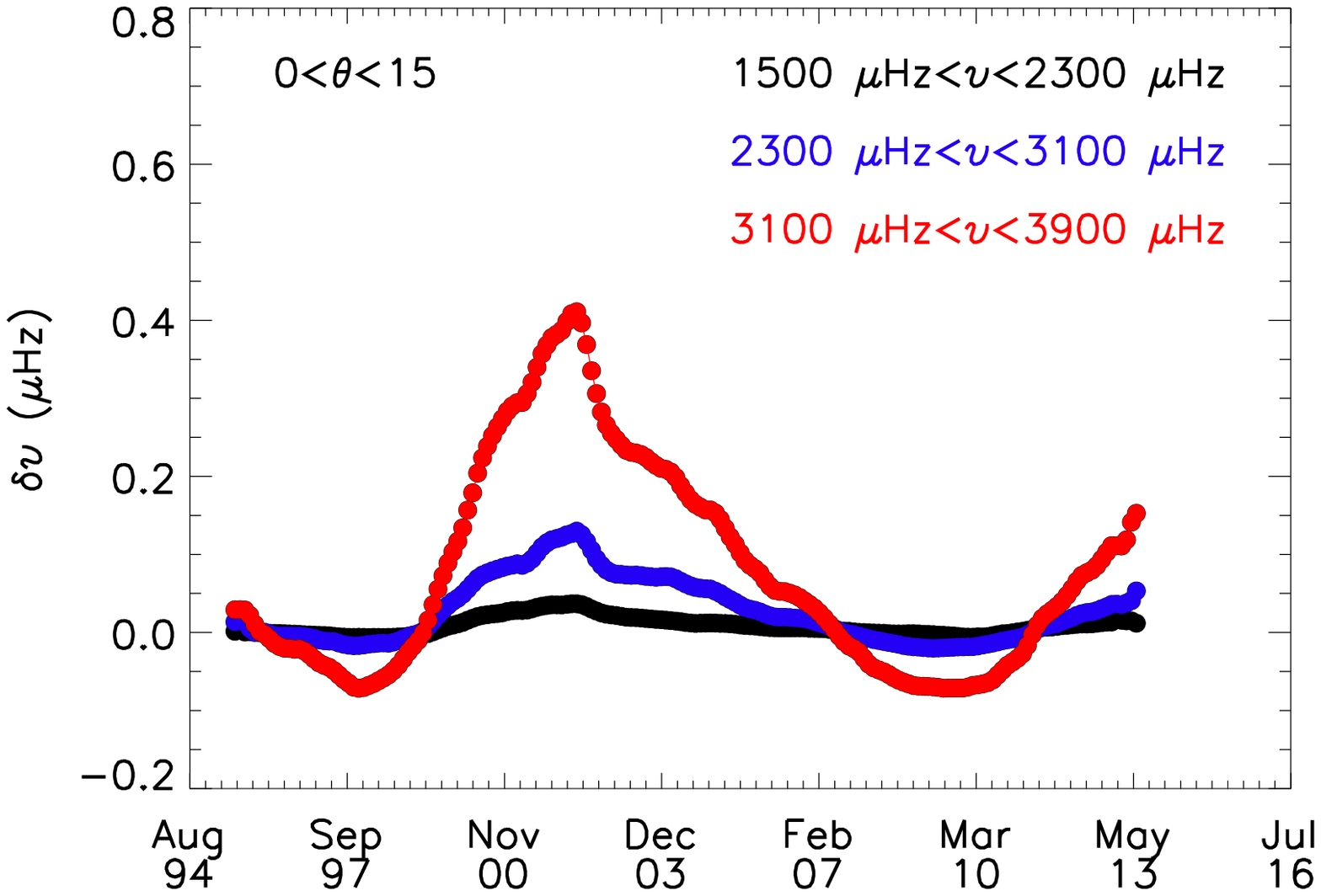}
\includegraphics[scale=.45]{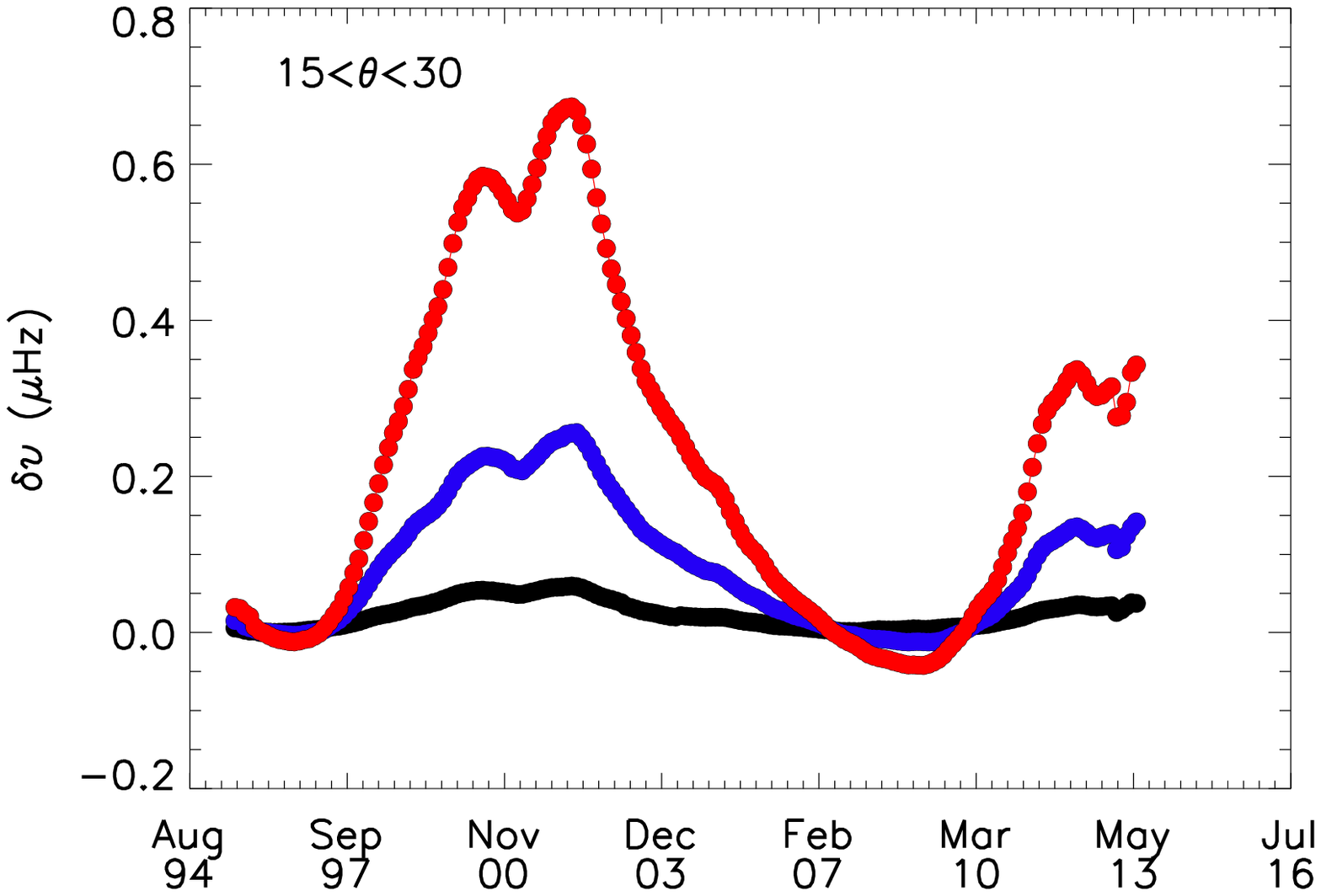}}
\centerline{
\includegraphics[scale=.45]{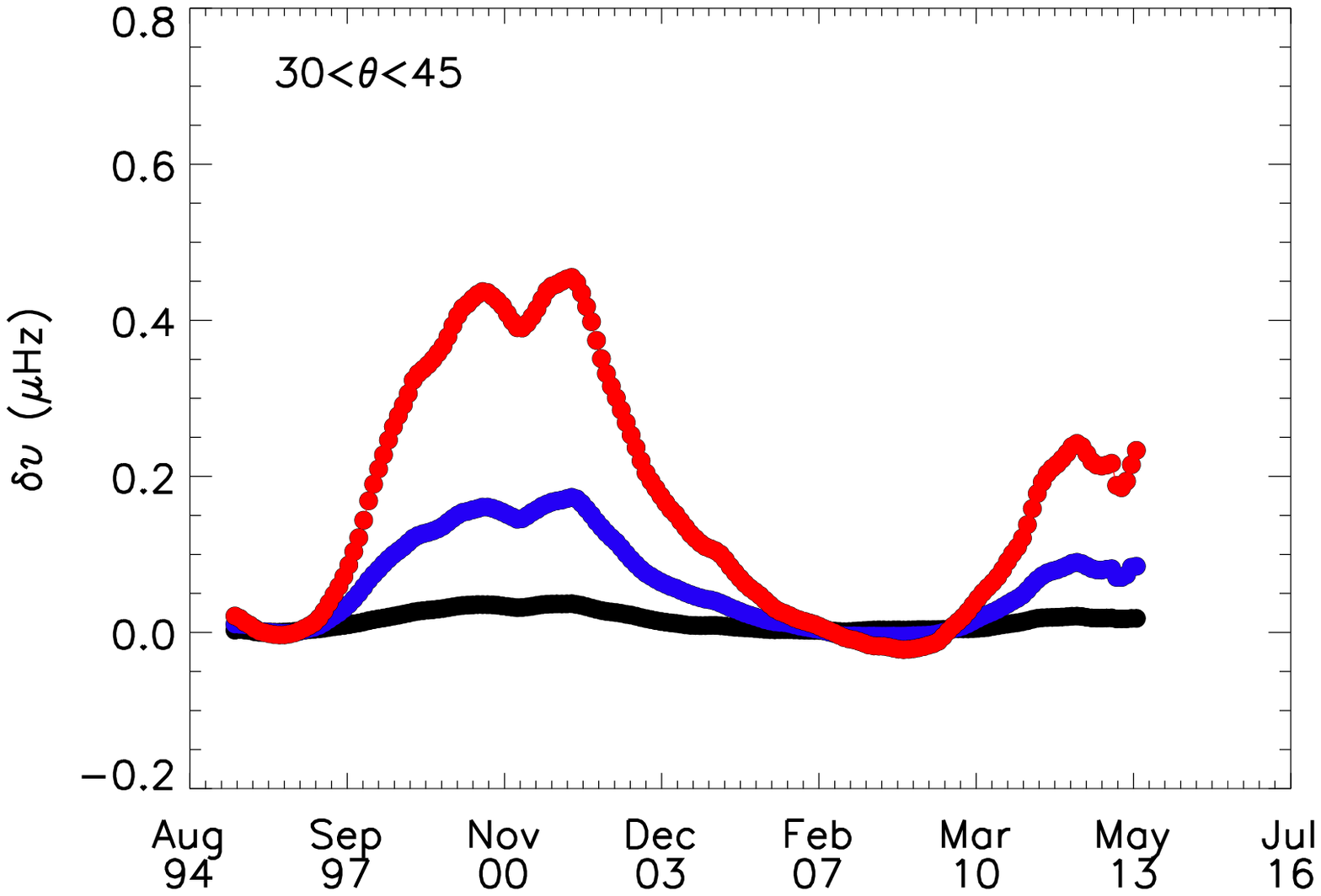}
\includegraphics[scale=.45]{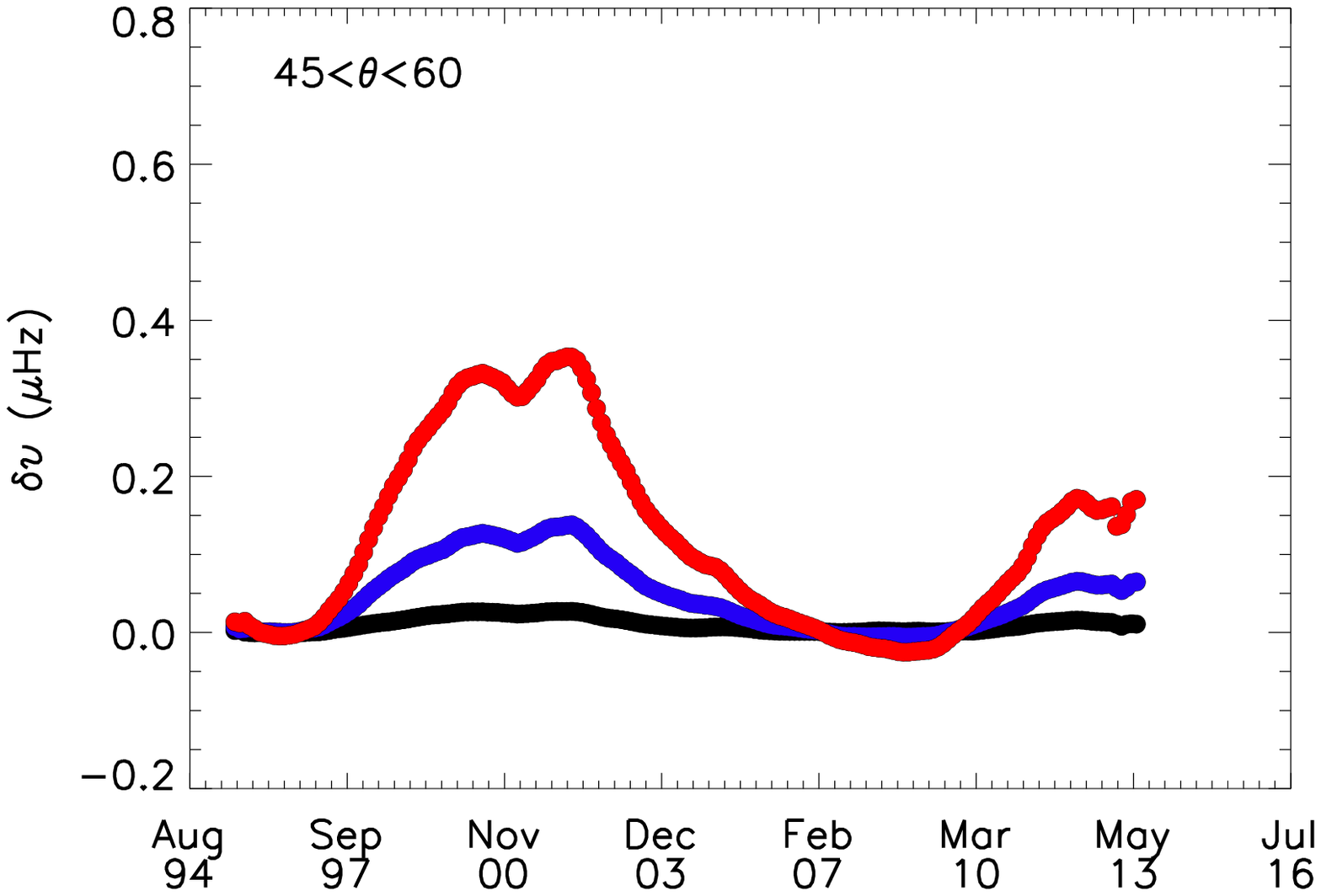}}
\centerline{\includegraphics[scale=.45]{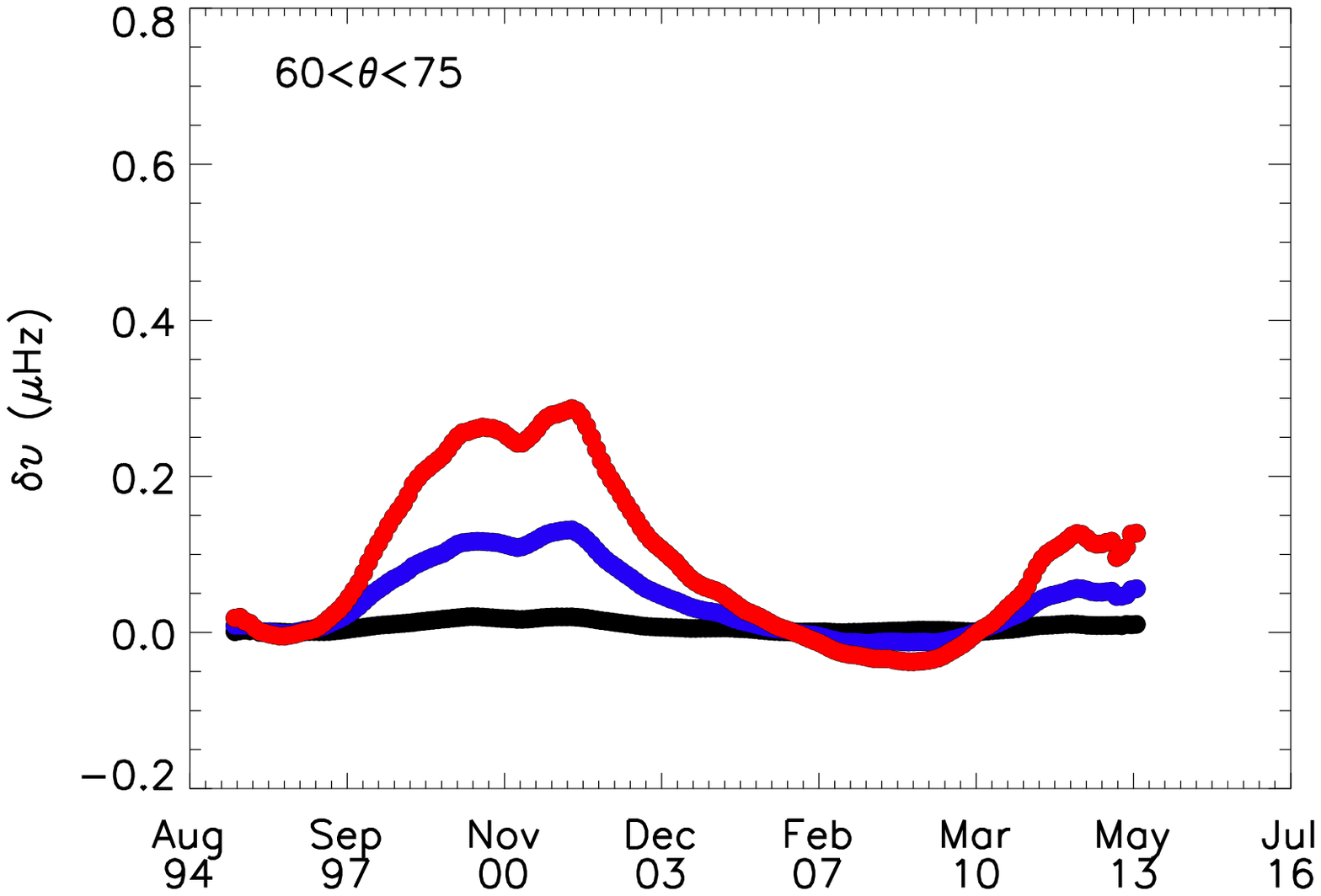}}
%\end{tabular}
\caption{Each panel shows the progression of solar cycle 23 and 24 at specific latitude but in three frequency bands corresponding to different subsurface layers. From top to bottom the progression of solar cycle with increasing latitude.}
\label{fig:difffreq_0_45}
\end{figure*}

\subsection{Progression of solar cycle in the Northern and Southern hemispheres}
{\textcolor{black}{The top two panels of Fig.~\ref{fig:high_hemis} show the progression of solar cycle in the Northern and Southern hemispheres as determined by the analysis of the high degree modes in two different latitudinal bands. While in the Northern hemisphere the activity is almost comparable at the two selected latitudes, the activity in the Southern is higher at the maximum between 15\degr\ $\le\theta\le$ 30\degr\ latitude, to then reach comparable strengths sometime after December 2003. After September 2006 the activity between 0\degr\ $\le\theta\le$ 15\degr\ and 15\degr\ $\le\theta\le$ 30\degr\ latitudes followed different patterns.  In fact, while below 15\degr\ we observe a prolonged minimum until June 2009 with a consequent delay in the onset of solar cycle either 
hemisphere (shown by black), above 15\degr\ latitude the rising phase already started sometime after September 2006 (shown by red).} 
%In fact, while below 15\degr\ we observe a prolonged minimum until June 2009 with a consequent delay in the onset of solar cycle either in the Northern 
%or in the Southern hemisphere, above 15\degr\ latitude the rising phase already started sometime after September 2006. 
\begin{figure*}
\centerline{
\includegraphics[scale=.7, trim=0 0 0 0]{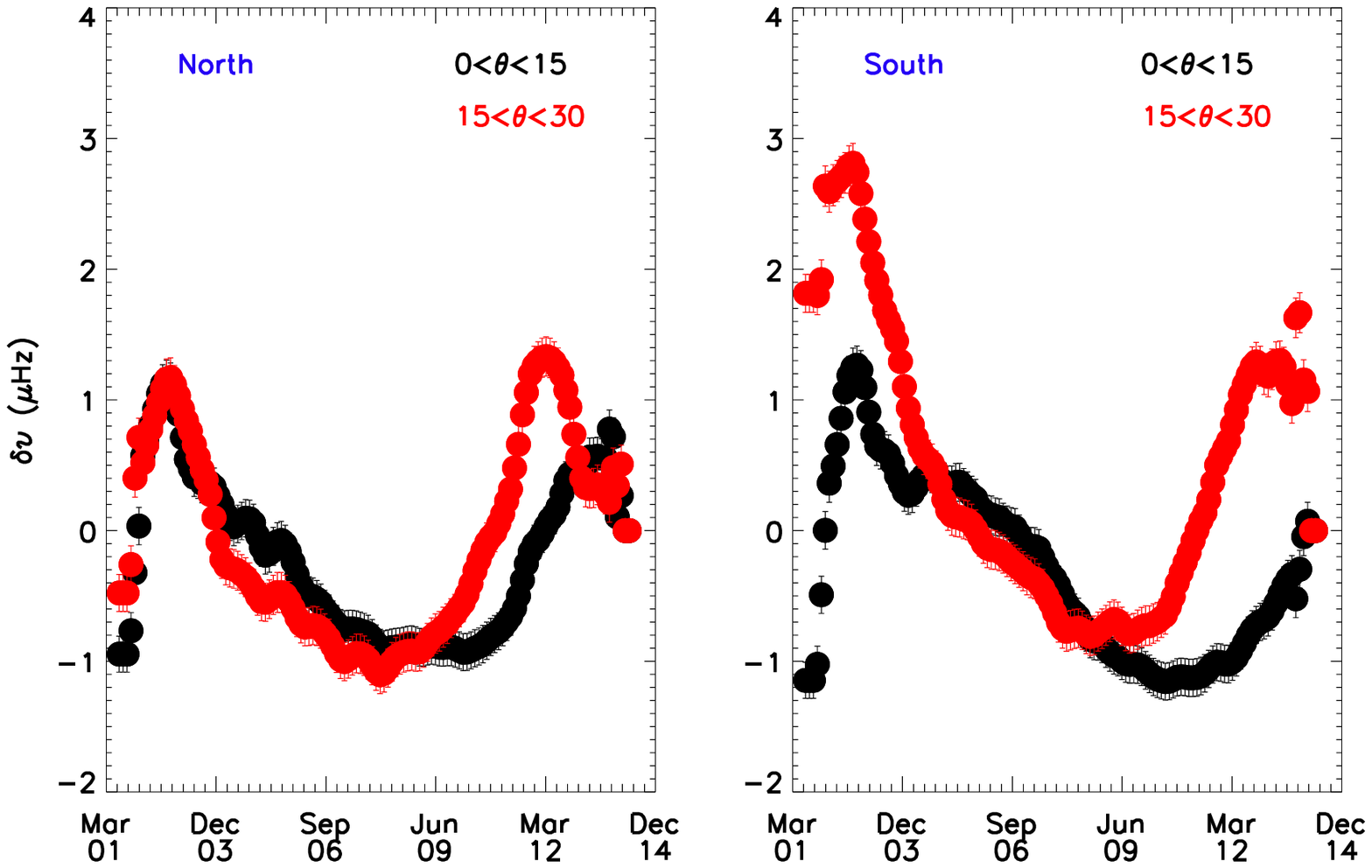}}
\centerline{
\includegraphics[scale=.7]{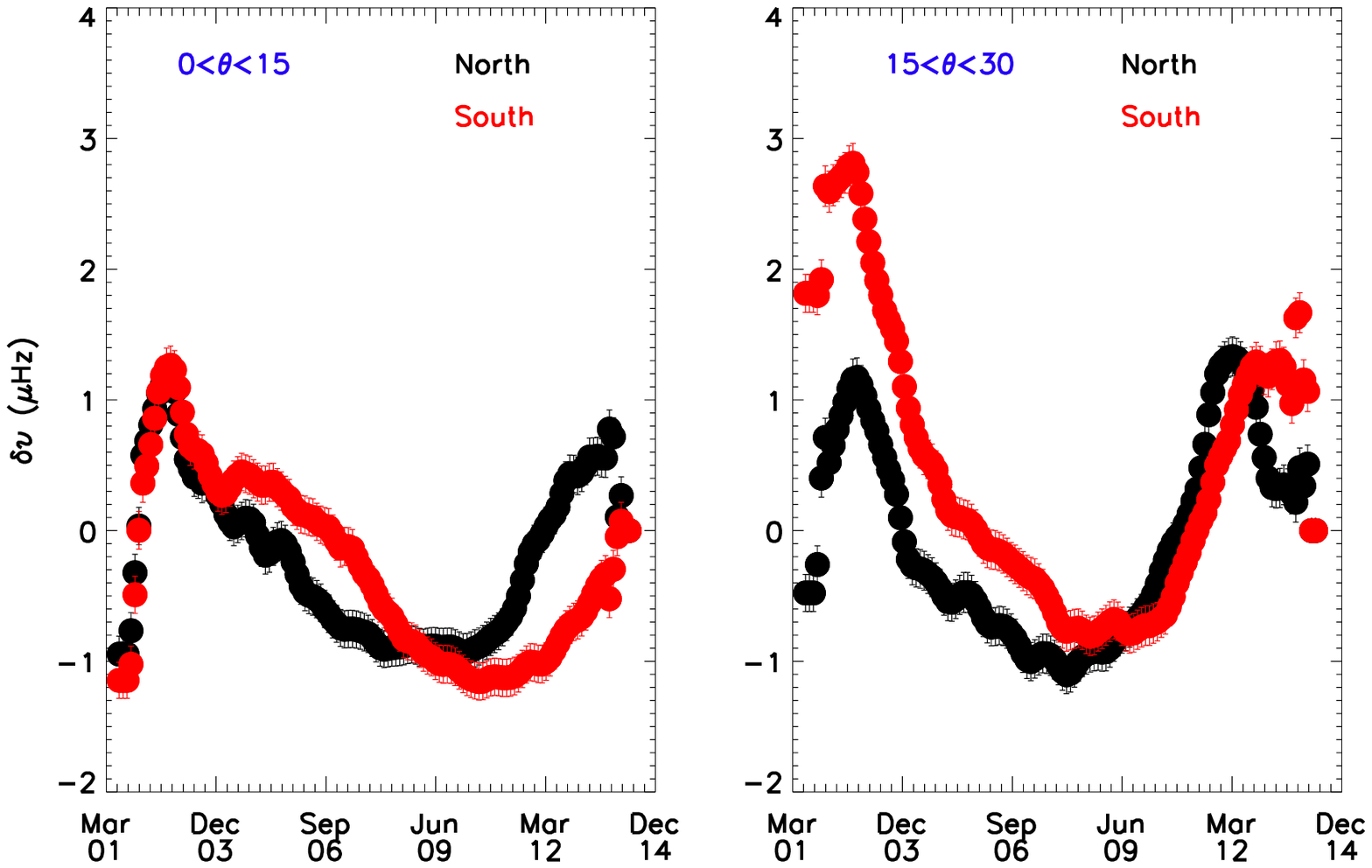}}
\caption{Top two panels: Progression of the solar cycle as seen in high-degree mode frequency shifts in northern (left) and southern (right)
hemispheres in two different latitude ranges. Bottom two panels: Comparison of the progression of solar cycle between northern and southern 
hemispheres in the same latitude range; 0\degr\ $\le\theta\le$ 15\degr\ (left), and 15\degr\ $\le\theta\le$ 30\degr\  (right)}
\label{fig:high_hemis}
\end{figure*}
When we compare the strength of activity at the same latitude but for different hemispheres (bottom two panels of Fig.~\ref{fig:high_hemis}), we find that throughout the 
descending phase the Southern hemisphere (shown by red) has been more active compared to the 
Northern one (shown by black). In particular we note an enhancement in the activity, 
which changes the evolutionary path of the descending phase at latitudes between 0\degr\ $\le\theta\le$ 15\degr\ (bottom left panel). This deviation is stronger in 
the Southern hemisphere. %Table 2 lists the time of the minimum in the two hemispheres and in two latitudinal bands. 
Interestingly where the excess of magnetic activity is more pronounced, the descending phase is consequently slightly more prolonged, leading 
to a longer overlap of successive cycles. Table 2 lists the time of the minimum in both Southern and Northern hemisphere and at two latitudes. 
In the Southern hemisphere the overlap of successive cycles lasted slightly more than  in the Northern hemisphere. 
Comparing the timing of the minimum between both hemispheres, we note that
the two hemispheres are in delay with respect to each other by approximately one year.

\begin{table}
\begin{center}
%\small\addtolength{\tabcolsep}{-3pt}
\caption{\textcolor{black}{The minimum epochs in the two hemispheres as seen from high-degree modes}}
%\small\addtolength{\tabcolsep}{-3pt}
\scalebox{1}{
\begin{tabular}{|c|c|c|c|}
\tableline\tableline
Latitude&\multicolumn{2}{c|}{Cycle 23}&Min-Min\\ \hline
Degree&Min South&Min North&Diff\\ \hline
0$\le\theta\le$15& 10/2010&09/2009&13 \\ \hline
15$\le\theta\le$30& 06/2008&09/2007& 9\\ \hline
\tableline
%\caption{The onset of solar cycle in the two hemispheres}
\end{tabular}}
\end{center}
\label{tab:minimum_hemis}
\end{table}
\section{Correlation with Sunspot}
\subsection{STARA data}
In order to compare the variation of oscillation frequencies with
known proxies of the solar activity, we use sunspot numbers calculated
from the Sunspot Tracking And Recognition Algorithm (STARA; \cite{Wat11}). In STARA sunspot catalogue, the sunspot count does not include
a factor for grouped sunspots and so the number is far lower than other
sources. The sunspot numbers are calculated using the MDI images for the period
from June 1996 to October 2010.  Although, there are some gaps in data
due to the SOHO vacation in 1998-99, the advantage of using this catalogue
over others is the availability of the location of sunspots on solar
disk which is important in this analysis. The sunspot numbers beyond
2010 are also available but have been calculated using the HMI images
which have different spatial resolution and no scaling has been performed
yet.
\begin{figure}[!b]
\center
\includegraphics[scale=0.42,trim=0 0 0 0]{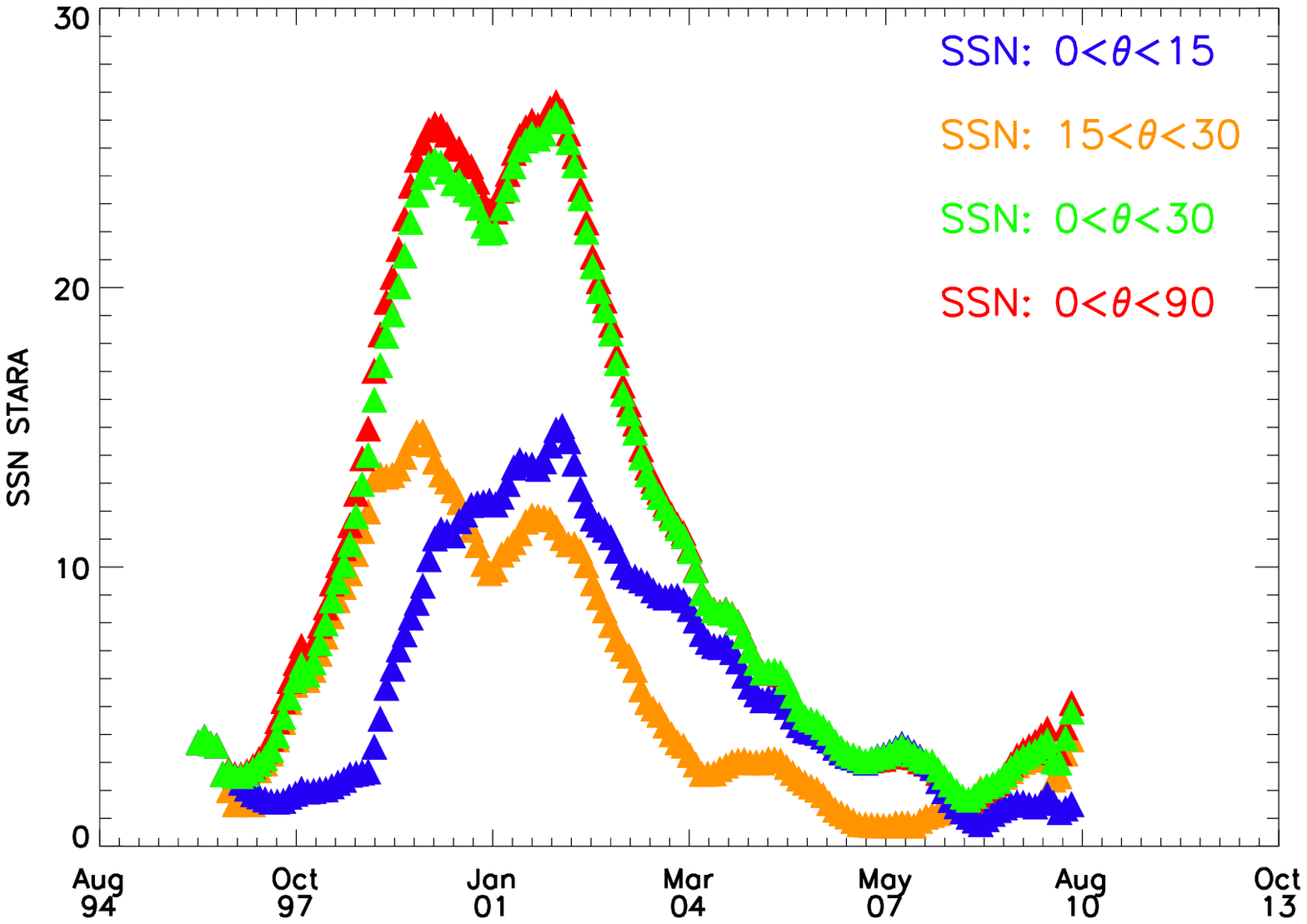}
\caption{The STARA Sunspot Number at different latitudes as indicated in the figure.}
\label{fig:stara}
\end{figure}
 Fig.~\ref{fig:stara} shows the sunspot numbers as measured by STARA; the gaps are due to unavailability of data. \textcolor{black}{The sunspot numbers are the averages over the same period as the time series of the oscillation frequencies. We further
 grouped them in three latitude bands. In the analysis of global modes, the selection of a particular latitude range does not confine modes' sensitivity to 
the selected range. Instead it senses all latitudes between northern and southern hemispheres up to the highest 
selected latitude. Thus frequency shifts in 15\degr\ $\le\theta\le$ 30\degr\ range, which covers modes in latitudes $\pm$ 30\degr\, are compared 
with the sunspot data between $\pm$ 30\degr\ latitude.
% In fact since the modes are sensitive to the entire latitude range less than the maximum value, the STARA data within 15\degr\ $\le\theta\le$ 30\degr\  does not include the contribution from 0\degr\ $\le\theta\le$ 15\degr\, while 0\degr\ $\le\theta\le$ 30\degr\. does include the range 0\degr\ $\le\theta\le$ 15\degr\ .
  There were no sunspots observed in 15\degr\ $\le\theta\le$ 30\degr\ latitude ranges
 before 1996 August, while some were visible in 0\degr\  $\le\theta\le$ 15\degr\ bands since the beginning of the available  data,
 i.e., 1996 June.}
  
Over the minimum phase between cycle 22 and 23, we notice that magnetic cycle started when the activity of the old cycle was still ongoing below 15\degr\ latitude. We also note that the maximum is characterized by a double peak structure for latitudes between 0\degr $\le\theta\le$ 30\degr, while by a single peak structure between 0\degr\  $\le\theta\le$ 15\degr , as already found in our helioseismic analysis of intermediate degree modes. 
 Furthermore soon after December 2003 an excess of sunspots at latitudes of 0\degr\ $\le\theta\le$ 15\degr changed the natural evolution of the activity during the descending phase. 
 
 \subsection{Frequency shifts and STARA Sunspots}
We aim at comparing STARA Sunspot number with helioseismic observations to highlight similarities and differences in the progression of the cycle between the two activity proxies. They are sensitive to different magnetic field structures. Sunspot are the result of the strong toroidal fields located at around equatorial latitudes, while acoustic waves sounds the whole Sun at all latitudes. Therefore they are sensitive to strong and weak toroidal fields. \textcolor{black}{To compare the behavior of the two activity proxies, we treated the STARA SSN data as we did for the mode frequency. We determined the SSN reference values (SSN$_{ref}$) between December 1996 - March 1997, the common period of quiet activity phase where SSN at all latitudes are around two. Then we calculate the deviation $\delta SSN$ as the differences between the observed sunspot at different epochs (SSN) and the SSN$_{ref}$. Finally $\delta SSN$ and $\delta\nu$ have been divided for the sum of the shifts/sunspots throughout the observational time, which gave us $\delta SSN_{rel}$ and $\delta\nu_{rel}$. 
Fig.~\ref{fig:shifts_ssn} compares the behavior of solar magnetic activity from STARA data with frequency shifts determined in the high frequency band at each selected latitude, as the high frequency range sounds the closest layer to the solar surface.
 It is worth reminding that, as the reference values for $\delta SSN$ and $\delta\nu$ have been calculated over different periods of activity and the length of observations span different time lengths, we cannot directly compare the size of $\delta SSN_{rel}$ and $\delta\nu_{rel}$, however the overall trend can be compared. Nevertheless we find that both activity proxies are characterized by a single peak structure at latitudes below 15\degr\, while above it by a double peak structure. This further confirms that the single peak structure is a signature of the solar magnetic activity at 0\degr\ $\le\theta\le$ 15\degr\ latitude ranges. }%{ We also note that, when we consider the 15\degr\ $\le\theta\le$ 30\degr\ latitude range, it is the first peak to be higher than the second one, while it is reversed when we consider 0\degr\ $\le\theta\le$ 30\degr\ .
% The latter fits better with $\delta\nu_{rel}$.}}
Although the origin of $p$-mode frequency shifts is 
still a matter of debate, magnetic fields in Sunspots can widen or shrink the acoustic cavity \citep{Sch06,Sim10}, shifting the mode frequency
towards lower/higher values. We might argue that during the descending phase, soon after December 2003, an excess of emergence 
of Sunspots produced the observed enhancement in the size of the shift predominantly in the Southern hemisphere and at latitudes below 15\degr\ .
\begin{figure*}
\centerline{
\includegraphics[scale=0.35, trim=0 0 0 0]{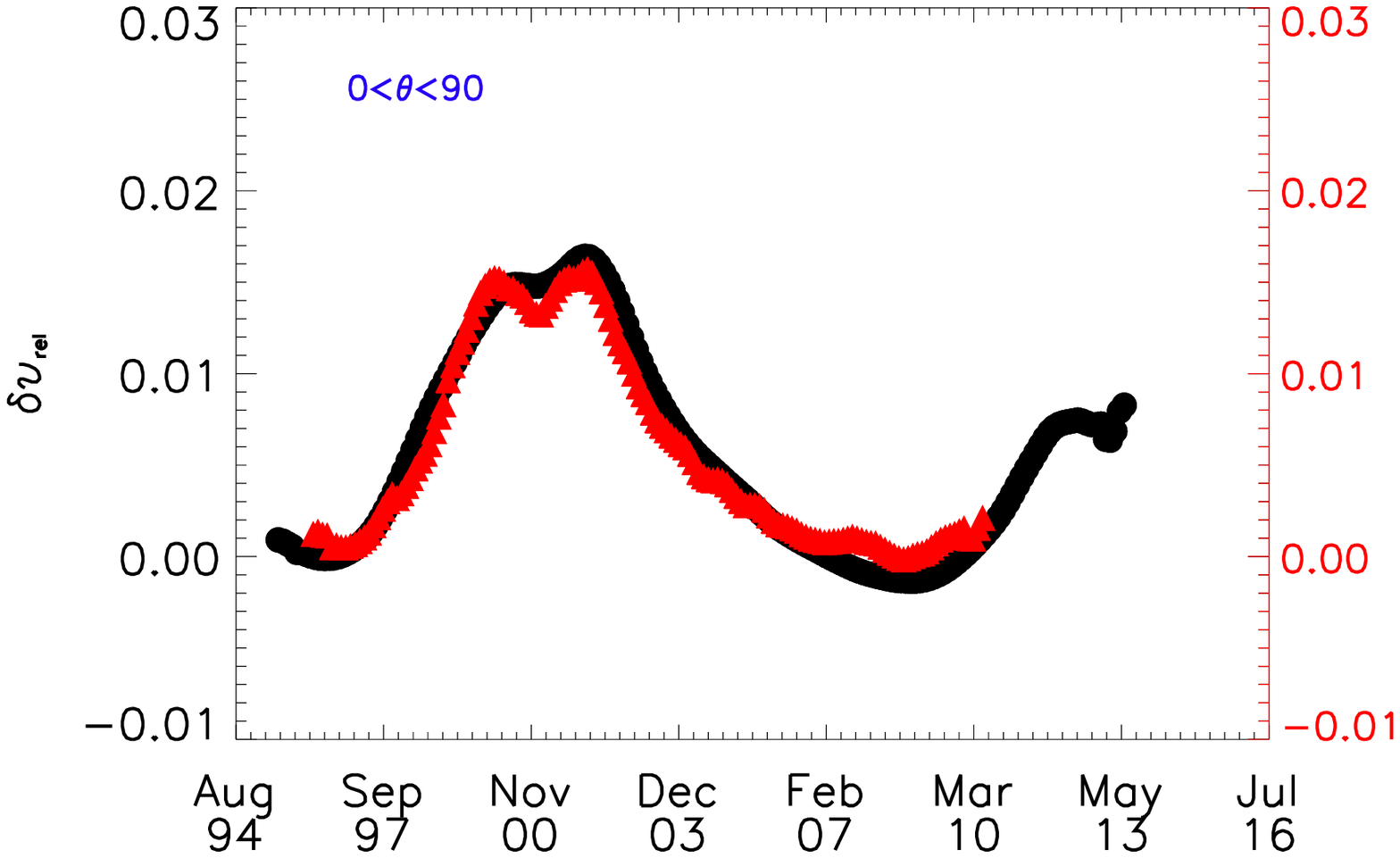}}
\centerline{
\includegraphics[scale=0.35]{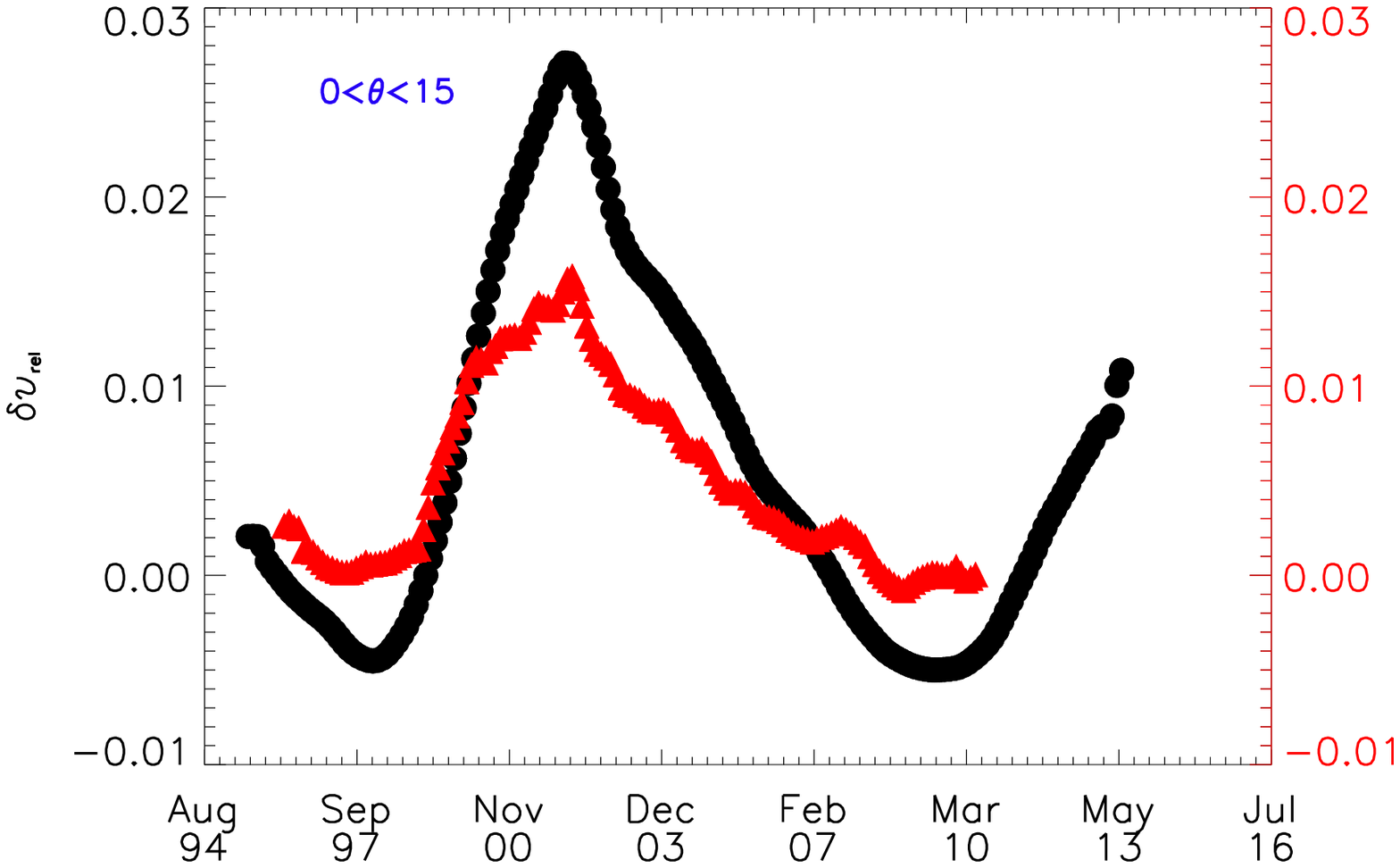}
\includegraphics[scale=0.35]{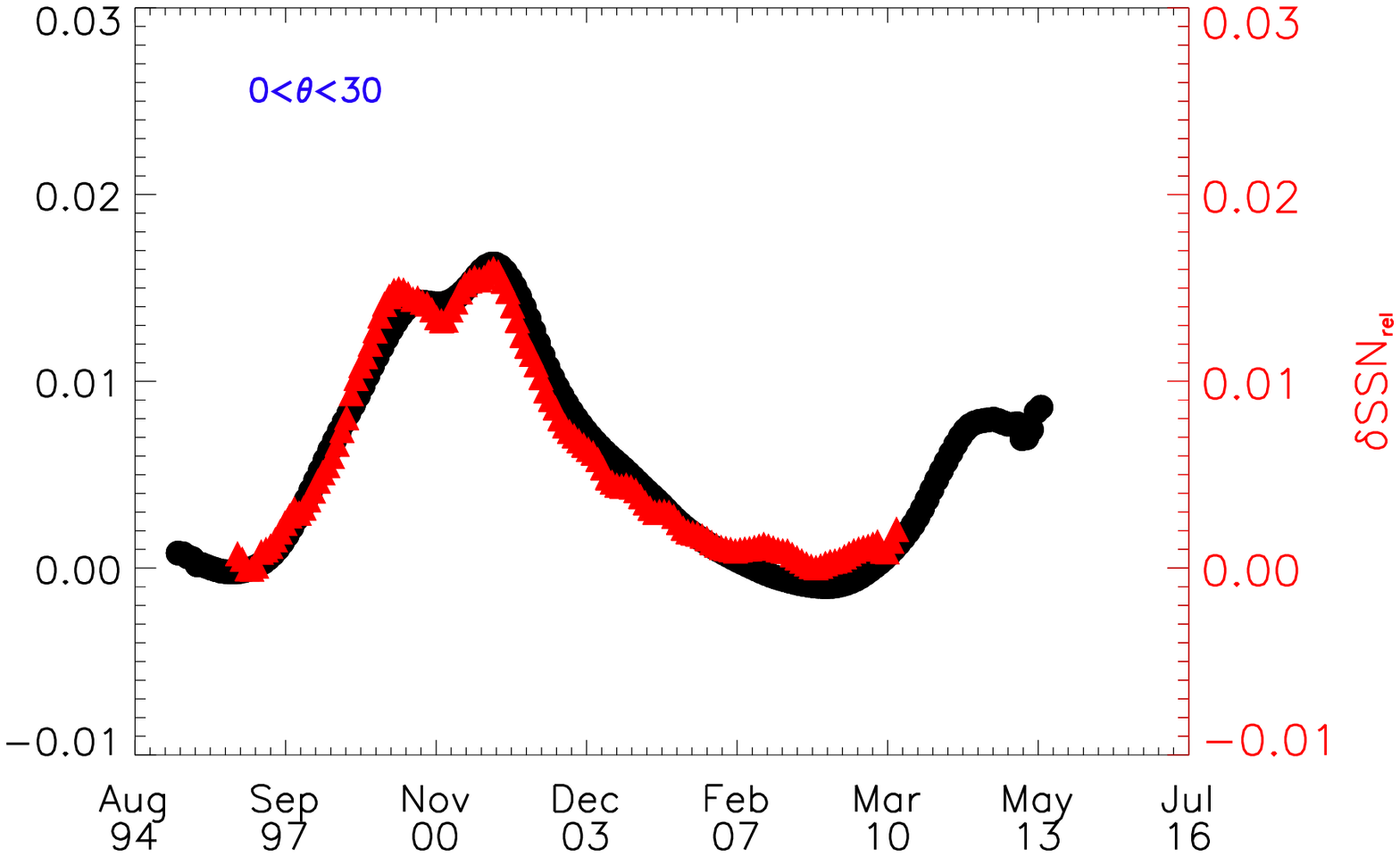}}
\caption{\textcolor{black}{Comparison between $\delta SSN_{rel}$  (red) and $\delta\nu_{rel}$ (black) at three different latitudes. Top panel
show the temporal variation of these two quantities  averaged over disk, while bottom row shows the variation in
two latitudinal bands: 0\degr\ $\le\theta\le$ 15\degr\ (left),  and 0\degr\ $\le\theta\le$ 30\degr\ (right).}}\label{fig:shifts_ssn}
\end{figure*}
\section{Discussion}
 Solar cycle changes in  p-mode frequencies are unique tools to investigate similarities and differences in the
progression of solar cycle at different latitudes and subsurface layers. In addition, high degree modes calculated
from local helioseismology techniques open a window on the solar hemispheric activity. In this work, therefore, we use
intermediate and high degree acoustic modes to obtain a detailed description of the SunÕs global and hemispheric
magnetism.

The latitudinal and frequency dependence of solar cycle changes in p-mode frequency shifts from intermediate degree
modes is a different representation of what is seen in the butterfly diagram and latitudinal inversions of the
helioseismic modes \citep{How02}. However, this approach highlighted new important details, which could be used
to constrain the sources of the $\alpha$ mechanism.

The overall results have pointed out latitudinal differences in the solar cycle progression below and above 15\degr\
in both hemispheres. The cycle onset below 15\degr\ is delayed compared to higher latitudes, causing an overlap of
successive cycle at the time of the minimum phase. To this regard the analysis of high degree modes identified and
quantified, for the first time, differences in the length of the overlap of successive cycles in the two hemispheres. { Furthermore
in both hemispheres our findings confine the overlap at latitudes below 15\degr\ .} 
Soon after the minimum, the activity level below 15\degr\ progresses with the fastest rise time and it reaches the
maximum characterized by a single peak. At higher latitudes instead the ascending phase is characterized by a slower
rise time and it ends up in a maximum characterized by a double peak structure. Interestingly the single peak below
15\degr\ coincides with the second and highest peak at higher latitudes. The dynamo mechanism seems to synchronize
the epoch of the maximum at all latitudes. During the descending phase we found latitudinal differences in the decay
time, as it is slower below 15\degr\ compared to higher latitudes.

How these observed properties can help us in understanding the role of the BL poloidal field sources with respect to
the  $\alpha$ turbulent one?{ For example, our findings have provided evidences that the overlap occur at latitudes below 15\degr\ . This confinement is better reproduced in FTD models including the Babcock - Leighton mechanism. Therefore we might envisage that to some extension the Babcock-Leighton mechanism might play a role in the solar dynamo \citep{Cam15}, but to draw any conclusion all the solar cycle features need to be reproduced within this formalism.} 
In fact the meridional flow speed sets the cycle period, the rise and decay time within BL flux
transport dynamo. Within the  $\alpha$ turbulent $\omega$ dynamo the cycle period depends (among others) on the
$\alpha$ turbulent effect itself \citep{Par55}. It would be then rather interesting to see how our findings will
impose further constraints on the latitudinal dependence of the meridional flow speed and $\alpha$ turbulent effect.
How well the resulting simulations will fit with observations, will be a valuable test to discern among the multitude
of dynamo models and it will shed some light on the principle driving solar dynamo.
\acknowledgements
We thank the anonymous referee for his critical comments. This work utilizes GONG data obtained by the NSO Integrated
Synoptic Program (NISP), managed by the National Solar Observatory, which is operated by AURA, Inc. under a
cooperative agreement with the National Science Foundation. The authors thank F. Watson for providing us the STARA
data.
	% (uses file "plain.bst")
\bibliography{biblio}
%\bibliography{massloss}
\end{document}